%% file: main.tex
\pdfoutput=1
\documentclass[conference]{IEEEtran}
\IEEEoverridecommandlockouts
\usepackage{cite}
\usepackage{amsmath,amssymb,amsfonts}
\usepackage{mathtools}
\usepackage{algorithmic}
\usepackage{graphicx}
\usepackage{textcomp}
\usepackage{hyperref}
\usepackage{comment}
\usepackage{cleveref}
\usepackage{footmisc}
\usepackage{multirow}
\usepackage[table,xcdraw]{xcolor}
\usepackage[ruled,vlined,linesnumbered,longend]{algorithm2e}
\usepackage{color}

\newcommand{\todo}[1]{}

\newcommand{\extended}[1]{}

\crefname{equation}{equation}{equations}
\Crefname{equation}{Equation}{Equations}
\crefrangelabelformat{equation}{(#3#1#4--#5#2#6)}

\crefmultiformat{equation}{equations (#2#1#3}{, #2#1#3)}{#2#1#3}{#2#1#3}
\Crefmultiformat{equation}{Equations (#2#1#3}{, #2#1#3)}{#2#1#3}{#2#1#3}

\newcommand{\topic}[1]{}

\def\BibTeX{{\rm B\kern-.05em{\sc i\kern-.025em b}\kern-.08em
    T\kern-.1667em\lower.7ex\hbox{E}\kern-.125emX}}
\begin{document}

\title{NoSQL Schema Design for Time-Dependent Workloads
}

\author{\IEEEauthorblockN{Yusuke Wakuta}
\IEEEauthorblockA{
%\textit{dept. name of organization (of Aff.)} \\
\textit{CyberAgent, Inc.}\\
Tokyo, Japan \\
wakuta\_yusuke@cyberagent.co.jp}
\and
\IEEEauthorblockN{Michael Mior}
\IEEEauthorblockA{\textit{Department of Computer Science} \\
\textit{Rochester Institute of Technology}\\
Rochester, NY, USA \\
mmior@mail.rit.edu}
\and
\IEEEauthorblockN{Teruyoshi Zenmyo}
\IEEEauthorblockA{
\textit{CyberAgent, Inc.}\\
Tokyo, Japan \\
zenmyo\_teruyoshi@cyberagent.co.jp}
\and
\IEEEauthorblockN{Yuya Sasaki}
\IEEEauthorblockA{\textit{Graduate School of Information Science and Technology} \\
\textit{Osaka University}\\
Osaka, Japan \\
sasaki@ist.osaka-u.ac.jp}
\and
\IEEEauthorblockN{Makoto Onizuka}
\IEEEauthorblockA{\textit{Graduate School of Information Science and Technology} \\
\textit{Osaka University}\\
Osaka, Japan \\
onizuka@ist.osaka-u.ac.jp}
}

\maketitle

\begin{abstract}
In this paper, we propose a schema optimization method for time-dependent workloads for NoSQL databases.
In our proposed method, we migrate schema according to changing workloads, and the estimated cost of execution and migration are formulated and minimized as a single integer linear programming problem.
Furthermore, we propose a method to reduce the number of optimization candidates by iterating over the time dimension abstraction and optimizing the workload while updating constraints.
\extended{
In addition, during migration candidate enumeration, we propose a method that reduces candidates by focusing on the similarity of query plans at each time and the key attributes of the schema.
Our evaluation shows that our proposed method achieved performance superior to static optimization in a variation of TPC-H with a time-varying workload.}
\end{abstract}


\section{Introduction}

\input{parts/1_Intro_eng}

\section{Preliminary}\label{sec:problem}

\input{./parts/2_ProblemStatement.tex}

\section{Proposed Optimization Formula}\label{Sec:Optimization}

\input{./parts/3_OptimizationFormulation}

\section{Proposed System}\label{sec:proposal}

\input{./parts/4_ProposedMethod}

\extended{
\section{Experimental Study}\label{sec:experiments}
\input{./parts/5_Experiments}
}

\section{Related Work} \label{sec:related}

\input{./parts/6_RelatedWork}

\section{Conclusion} \label{sec:conclusion}
We proposed new techniques for optimizing time-series schema by effectively reducing the number of schema candidates and the number of migration plan candidates.
First, we formulated the optimization problem of time-series schema with a single integer linear program for minimizing the total cost of time-dependent workload execution and database migration. 
Second, we proposed an efficient schema candidate pruning technique by decomposing the ILP of  original time-dependent workload via approximation into hierarchical local ILPs of smaller sub-workloads. This technique is scalable by effectively pruning uninteresting schema candidates.
Finally, we proposed an effective technique that reduces the number of migration plan candidates by restricting the candidates according to how optimized query plans are changed.
\extended{
既存手法の静的なスキーマ最適化手法では，時刻変化するワークロードの変化に追従できず，性能が低下する場合でも，提案手法により安定して高い性能を達成できることを評価実験を通して確認した．
As future work, we can consider optimizing for free time widths because the width between times is fixed, and optimizing when the frequency is unknown by combining with the query frequency prediction method ~\cite{Ma2018}.
}
%今後の課題として，時刻間の幅が固定であるため自由な時刻幅に対応した最適化やクエリ頻度の予測手法~\cite{Ma2018}と組み合わせて頻度が未知の場合での最適化が考えられる．

\section*{Acknowledgment}
This paper is based on results obtained from a project,
JPNP16007, subsidized by the New Energy and Industrial Technology Development
Organization (NEDO). 

\bibliographystyle{abbrv}
\bibliography{icde}  

\end{document}

%% file: parts/1_Intro_eng.tex
\label{sec:intro}
%%
%\memo{Background}
%%
Frameworks for big data management are widely used in many applications, such as Web services, IoT applications, and scientific analysis.
These applications need to manage petabyte-scale data.
In particular, NoSQL databases are one of the important frameworks for such large-scale big data management.
Historically, many NoSQL databases have their roots in systems such as BigTable~\cite{DBLP:journals/tocs/ChangDGHWBCFG08}, Amazon Dynamo~\cite{DBLP:conf/sosp/DeCandiaHJKLPSVV07}, and Yahoo! PNUTS~\cite{DBLP:journals/pvldb/CooperRSSBJPWY08}. Some recent examples of NoSQL databases are Google F1~\cite{DBLP:conf/sigmod/ShuteOEHRSVWCJLT12} and Spanner~\cite{DBLP:journals/tocs/CorbettDEFFFGGHHHKKLLMMNQRRSSTWW13}: most of these systems are classified as wide-column store (or extensible record store), a general type of NoSQL databases.

%%
%\memo{Background: clarify the importance of predictable cyclic patterns on workload}
%%
In most of the above applications, workloads follow common time-dependent patterns, such as cycles, growth and spikes, or workload evolution~\cite{Ma2018}.
We focus on predictable or pre-scheduled patterns that are common features in automated IoT applications and scientific analysis.
Specifically, we describe two such examples:
\begin{itemize}
\item {\bf IoT applications.} An electric power company utilizes an analytic pipeline to collect power usage from 7.5 million smart meters, put them into a distributed database system, aggregate the power usage, compute its total cost, and then notify electrical power retailers in a timely fashion~\cite{Sasaki2017}.
\item {\bf Astronomical data analysis.} Astronomers use an analytic pipeline to capture images, transform data, calibrate parameters, identify objects, and detect transients and variables~\cite{Takata2020}.
The National Astronomical Observatory of Japan provides a database service on the Web to support such pipelines for astronomers.
The database stores 436 million objects and works on a distributed database system~\cite{PDR2}.
A large number of different queries are executed on the database depending on the type of analysis tasks. 
\end{itemize}
% Toward fast search and real-time inputs of big astronomical catalogs by the new generation relational database, https://www.adass2019.nl/wp-content/uploads/adass-pdf/Poster391.pdf
% Innovative astronomical applications with a new-generation relational database
% https://ui.adsabs.harvard.edu/abs/2020SPIE11452E..26F/abstract
An important feature of those automated analysis pipelines is that they form predictable patterns: workloads are scheduled in advance and they are repeated for a certain time period, such as every day or every week.
So, we can significantly improve the performance of database systems by appropriately designing database schema based on such predictable patterns.

\subsubsection*{Technical trend and Major issues}
Schema design on NoSQL is crucial for achieving high performance for large-scale big data management.
There has been significant research on automated schema design on relational databases~\cite{Agrawal2000,Kimura2010,Rafi2020,DBLP:journals/pvldb/TangSEKF20}, NoSQL~\cite{Vajk2013,Mior2017, CONST2020}, and cloud-scale environment~\cite{Jindal2018,DBLP:conf/socc/JindalPRQYSK19}.
Since our target is a large-scale big data management, we focus on the technical trend of schema design techniques for NoSQL (see Section~\ref{sec:related} for relational databases).
Most existing work assumes that the workload is static (does not change dynamically).
That is, they use an average workload aggregated from a dynamic time-dependent workload. 
As an example, NoSQL Schema Evaluator (NoSE)~\cite{Mior2017} leverages integer linear programming (ILP) for schema design to optimize the execution cost of static workload. 
%% The experiments validated the schema suggested by NoSE outperformed the design manually optimized by database experts. 
However, NoSE is not effective for a time-dependent workload because its average workload may not be a good approximation for the whole workload. 
Also, there is another type of research that adaptively changes schemas as workload changes.
CONST~\cite{CONST2020} is one such example, which heuristically changes schema design over time.
However, it ignores the cost of database migration when schema changes so it does not generate optimal schema designs for time-depended workloads.

\subsubsection*{Technical challenges}
To tackle the above weaknesses of existing techniques, we take an approach for optimizing time-series schema of taking both the execution cost of time-dependent workload and database migrations into account.
However, if we naively extend existing techniques designed for static workloads to dynamic time-dependent ones, we have three obstacles for finding optimal answers using ILP: 1) we need to handle the trade-off between the cost of time-dependent workload execution and database migration, 2) the number of schema candidates blows up depending on the number of time steps, and 3) the number of migration plan candidates also blows up depending on the number of schema candidates.
Here, a migration plan indicates a database transformation from old schema to new schema.

\subsubsection*{Contributions}
We propose new techniques for optimizing time-series schema by effectively reducing the number of schema candidates and the number of migration plan candidates.
The novelty of our proposal is three-fold.

%\memo{Proposal: 1st novelty}
First, we formulate the optimization problem of time-series schema with a single integer linear program for minimizing the total cost of time-dependent workload execution and database migration. 
%The formulation with a single integer linear programming is indispensable, since we need to handle the trade-off between the cost of time-dependent workload and database migrations.
%There is a trade-off between cost of time-dependent workload and database migrations that choosing the best schema at every time step require additional cost of database migration.
% Therefore, we formulate the optimization problem as single integer linear programming to consider the trade-off.
%\memo{Proposal: 2nd novelty}
Second, we propose an efficient schema candidate pruning technique by introducing a new data structure which we call a workload summary tree. % that approximates the ILP for original time-dependent workload.
We decompose the ILP of the original time-dependent workload via approximation into hierarchical local ILPs of smaller sub-workloads. This technique is scalable by effectively pruning uninteresting schema candidates, since each local ILP works efficiently with a small number of time steps while capturing the global feature of its parent workload.
%We design the workload summary tree to have the following features, 1) every child node represents a sub-workload split in time-scale from the workload of its parent node and 2) parent node's workload is more summarized than its child's sub-workload.
\begin{comment}
In detail, 
we recursively construct ILP for a summarized workload assigned to each node starting from the root node and translate its answer as the ILP constraint of its child sub-workloads in order to take the global features into account at local ILPs.
We identify uninteresting schema candidates that are never chosen as ILP answers at any nodes in the tree.
%This approach is efficient, because the time step size of (sub-)workload at every node is kept significantly smaller than that of the original workload.
\end{comment}
%\memo{Proposal: 3rd novelty}
Finally, we propose an effective technique that reduces the number of migration plan candidates.
We notice that workload changes cause optimized query plan changes, which necessitate a database migration. 
So, we can effectively reduce the number of migration plan candidates by restricting them according to how optimized query plans are changed instead of considering arbitrary migration plans.
% among possible query plan candidates. 
\begin{comment}
We first enumerate migration plan candidates query-wisely by ignoring the correlation between queries.
Then, we enumerate additional migration plan candidates that capture the correlation between queries, but we limit only to efficient ones: the key order in an old schema (partition key and clustering key in column families) is kept in the new schema after the migration.
\end{comment}
%We validate the effectiveness of our proposal through intensive experiments by adapting the TPC-H benchmark to generate a time-dependent workload.

\subsubsection*{Paper organization}
The rest of this paper is organized as follows.
We describe the problem statement and the challenges of schema design problem for time-dependent workload (Section~\ref{sec:problem}) and then formulate it with a single integer linear program (Section~\ref{Sec:Optimization}).
We describe the detail of our approach %including a system overview, a column family pruning using workload summary tree, migration plan enumeration, and cost estimation 
(Section~\ref{sec:proposal}). 
\extended{
We validate the effectiveness of our approach through intensive experiments using an extended TPC-H benchmark to various time-dependent workloads (Section~\ref{sec:experiments}).}
We finally position our proposal with respect to the state of the art (Section~\ref{sec:related}) and give concluding remarks (Section~\ref{sec:conclusion}).

%% file: parts/2_ProblemStatement.tex
In this section, we describe a schema optimization problem for time-dependent workloads on NoSQL databases, in particular on extensible record stores~\cite{Cattell2011}.

\subsection{Extensible Record Store}\label{ssec:ERS}
An extensible record store, such as Apache Cassandra, is a general type of NoSQL databases that achieves high scalability by partitioning database in distributed multiple nodes. 
Extensible record stores utilize column families (CFs) for expressing database schema.
We treat CFs as physical schema, which is derived from conceptual schema expressed with an entity graph~\cite{Mior2017} (simplified ER model).
CFs contain three types of columns, \emph{partition  key}, \emph{clustering  key}, and \emph{value}. 
\emph{partition key} is a key used for database (range or hash) partitioning over multiple nodes. 
$clustering \: key$ is a sorting key used inside for each node. 
Other columns are treated as $values$.
CF is expressed in the following notation:
\begin{equation}
  [\mathit{partition \: keys}][\mathit{clustering \: keys}] \rightarrow [\mathit{values}] \nonumber
 \label{Equation:CF}
\end{equation}
The notation indicates a functional dependency from (\emph{partition keys}, \emph{clustering keys}) pair to \emph{values}.

\subsection{Time-dependent Workloads}\label{ssec:TDWorkload}
We focus on predictable or pre-scheduled time-dependent workload patterns, such as cycles, growth and spikes, and workload evolution~\cite{Ma2018}, 
which are common patterns in automated IoT applications and scientific analysis.
In such workloads, all the queries and update operations are often known or predictable beforehand.
A time-depended workload $W$ denotes a collection of SQL statements, queries $Q$ and update operations $U$, on a conceptual (relational) schema
and only the frequency of each query/update operation changes over time\footnote{A new query appears if its frequency is changed from zero.}. 
Since the frequencies of queries/update change, we may need database migrations for reducing the total cost of the workload execution and the migration. 
%Since the frequencies change over time, an optimum physical schema (column families) at different time step can be different from those at other time steps. 
For example, a more denormalized schema should be chosen when the workload is read-intensive. 
This motivates us to optimize time-series physical schema over time-dependent workloads on extensible record stores.

\subsection{Problem statement}
Given a time-dependent workload, a conceptual schema, and a maximum storage size as a constraint, we identify an optimized time-series physical schema (set of column families) by minimizing the total cost of the workload execution and database migration.
As a result of schema optimization, we also output optimized query plans using the physical schema at each time step and migration plans between the schemas of different time steps.
A query plan is expressed on a physical schema, which is generated from a SQL statement in a workload $W$: it consists of multiple steps using filtering, sorting, and aggregation.
A migration plan transforms old column families into a new column family, which is generated from a migration query. Notice that we generate migration queries from workload queries $Q$, since database migration is usually made for reducing query cost by materializing query results (See migration plan enumeration in Section~\ref{Section:EnumerateMirgatePlan} for more detail).
In addition, new column families are incrementally maintained and workload continues execution during database migration. 

\subsection{Query/migration plan group}\label{ssec:QueryPlanAndMigrationPlan}
\begin{figure}[t]
  \begin{center}
    \includegraphics[scale = 0.65,bb = 0 0 382.08 315.12]{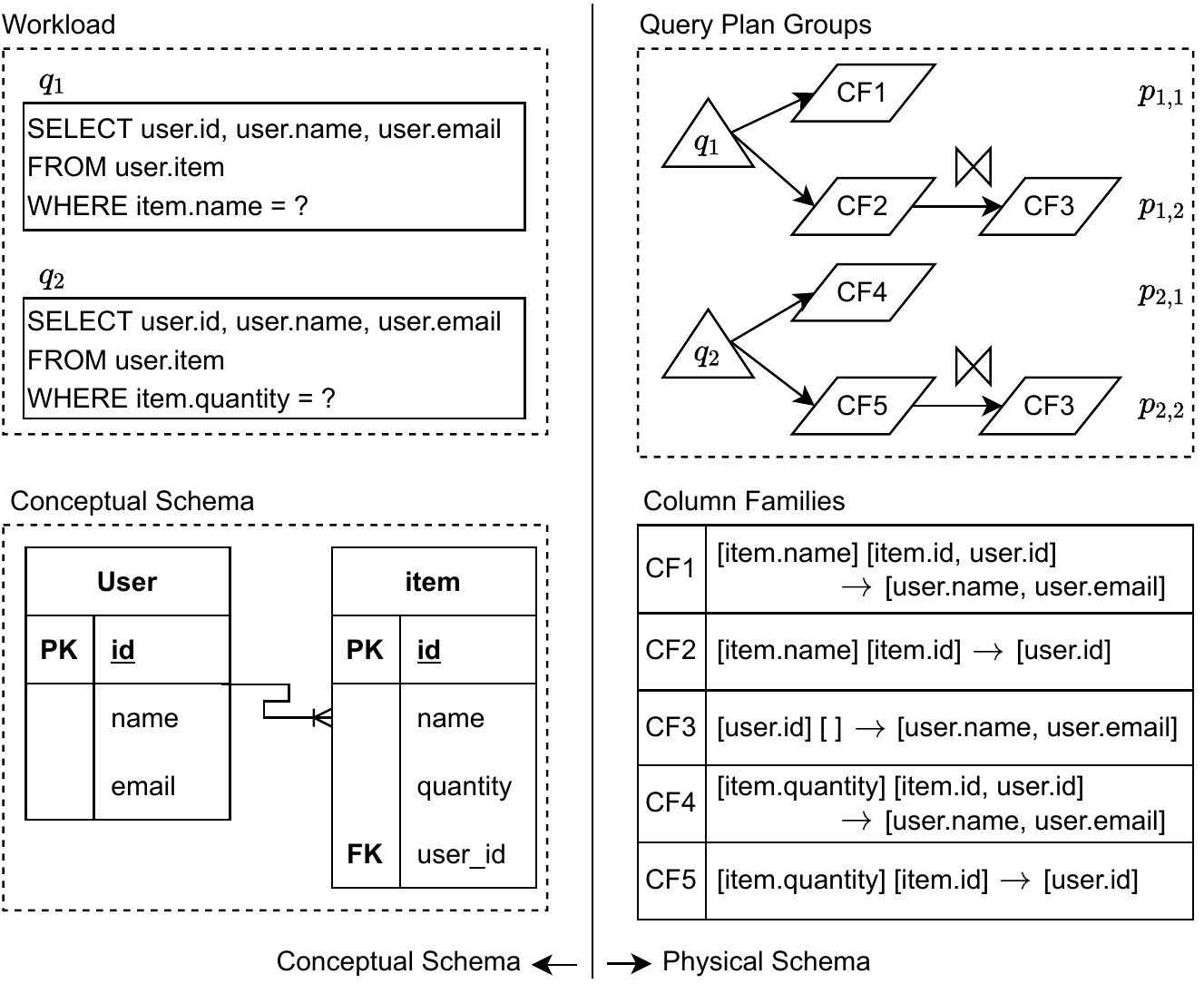}
    \caption{An example of schema design obtained from a conceptual schema for two queries, $q_1, q_2$.
A query plan group is generated on the column families enumerated from the columns used in each query.}
    \label{SchemaDesignExample}
  \end{center}
\end{figure}

\begin{figure}[t]
    \includegraphics[scale = 0.51,bb = 0 0 497 103.92]{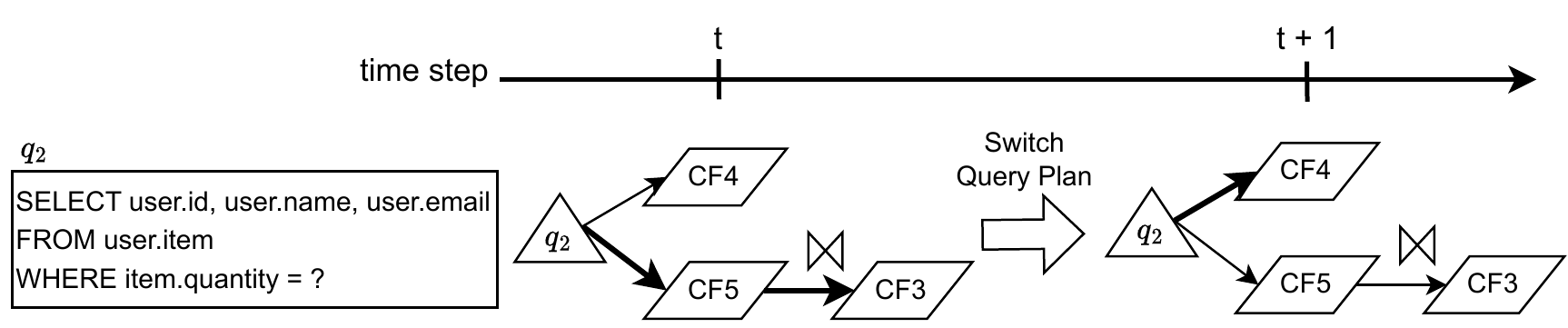}
    \caption{An example of changing query plans (the orange arrows indicate the chosen plan). The workload execution cost is reduced by changing the query plan according to the query frequency changes at each time step.\todo{crop}}
    \label{TdQueryPlanExample}
\end{figure}

We introduce query plan groups, a collection of query plans that are transformed from each SQL query in the workload\footnote{We similarly treat update operations as queries in the workload.}.
We choose a single optimized query plan in a query plan group at every time step by schema optimization.
We express a query plan as a path of steps (serialized from a typical query plan tree) where each step represents an operation, \texttt{Get} which fetches data from a column family. We call these steps \texttt{Get} steps for simplicity. 
Query plans in prior work~\cite{Mior2017} not only have \texttt{Get} steps but also has an \texttt{ORDER BY} step on column families at the server side additionally with join/filtering/sort steps that are executed at the application side.
We also express a query plan group using a tree structure which nodes with the same parent share a prefix of their query plans.
% Each query plan in a query plan group forms a path from the root node with following steps (parallelograms in the figure), where each step represents a get operation on its column family. 
We call a query plan with a single \texttt{Get} step a \emph{materialized view (MV) plan} and also call one with multiple \texttt{Get} steps a \emph{join plan}.
MV plans are efficient for query processing since they don't need join operations between column families.
In contrast, join plans use multiple normalized column families so they are efficient for update processing and saving storage size. However, they require expensive join operations between the column families, since extensible record stores do not support join operations at the server side so the join operations need to be made at the application side.

Next, we introduce migration plan groups, a collection of migration plans that are transformed from a migration query.
A migration plan generates a new column family at the next time step using schema at the current time step.
We choose a single optimized migration plan among multiple migration plan groups obtained from migration queries by schema optimization.

\subsection{Examples of optimizing time-series schema}
Figure~\ref{SchemaDesignExample} depicts an example of schema design obtained from a conceptual schema for two queries, $q_1, q_2$.
A query plan group is generated on the column families enumerated from the columns used in each query.
In Figure~\ref{SchemaDesignExample}, $p_{1, 1}$ and $p_{2, 1}$ are MV plans, and $p_{1, 2}$ and $p_{2, 2}$ are join plans.
For example, query plan $p_{1, 2}$ first extracts records from $\mathit{CF}2$ using an equality predicate on $item.name$ in query $q1$ and then extracts records from $\mathit{CF}3$ using $user.id$ as the join key from $\mathit{CF}2$ to $\mathit{CF}3$. 
Since join operations need to be made at the application side, the join plan $p_{1,2}$ is significantly slower than the MV plan $p_{1,1}$.
However, the join plan $p_{1, 2}$ requires less storage size than the MV plan $p_{1,1}$, since $p_{1, 2}$ uses denormalized schema, $\mathit{CF}2$, $\mathit{CF}3$.

Next, Figure~\ref{TdQueryPlanExample} depicts an example of changing query plans for a time-dependent workload. 
We assume that 
1) MV plans cannot be chosen both for $q_1$ and $q_2$ because of insufficient storage size, and 
2) $q_1$'s frequency is larger than $q_2$'s at time $t$ and they are reversed at time $t+1$; $q_1$'s frequency becomes smaller than $q_2$'s.
In this case, a join plan is chosen for $q_2$ at time $t$ and it is switched to a MV plan at time $t+1$ due to the query frequency change.
Such query plan changes reduce the workload execution cost at every time step, however they require the additional cost of database migrations.
Therefore, we need to choose an optimized time-series schema by considering the trade-off between workload execution cost and migration cost.
% 頻繁にマイグレーションを実行するとマイグレーションコストが増加するため，ワークロードのコストとマイグレーションのコストのトレードオフを踏まえたスキーマ設計が重要である．

%% file: parts/3_OptimizationFormulation.tex
We formulate the problem of time-series workload optimization using a single integer linear program (ILP) and output an optimized time-series schema, query plans at every time step, and migration plans between adjacent time steps.
The benefit of this approach is that it formulates the total cost of time-dependent workload execution and database migration using a single ILP, so it can handle the trade-off between the cost of time-dependent workload and database migration.
%the dependencies between query plans and migration plans are naturally handled in the single ILP.
%That is, we can not optimize the workload cost (query plans) and migration cost independently but we need to optimize both at the same time,
%because migration plans depend on the current schema at time $t$ and next schema at time $t+1$ and also query plans depend on the current schema, which is generated from migration plans.

For a given time-dependent workload consisting of queries $Q$ and update operations $U$, 
we obtain an optimized time-series schema $(S_1, ..., S_T)$ from time step $t = 1$ to $T$ by minimizing the following objective function using three constraints for query plans, migration plans, and storage size.

\noindent\\
{\bf{Objective function:}}
The objective of optimizing time-series physical schema ($S_1,\ldots, S_T$) is to minimize the total cost of time-dependent workload execution and database migration.
\begin{equation}
        \min_{S_1,\ldots, S_T} \; \sum_{t = 1}^{T} \mathit{workload}(S_{t}) + \sum_{t = 1}^{T - 1}\mathit{migrate}(S_{t}, S_{t + 1})\label{baseObjective:Objective}
\end{equation}
where $\mathit{workload}(S_{t})$ indicates the workload execution cost on schema  $S_{t}$ at time step $t$, and
$\mathit{migrate}(S_{t}, S_{t+1})$ indicates the migration cost from schema $S_{t}$ to $S_{t + 1}$.
If there is no migration ($S_{t}$ = $S_{t+1}$), then $\mathit{migrate}(S_{t}, S_{t + 1})$ = 0.

\subsection{Workload execution cost}
Workload execution cost is defined as the total cost of all queries and update operations in the workload;
each query/update operation cost is computed as the product of its frequency and its estimated execution cost. 
In detail, we define the workload execution cost at time step $t$ as follows:
\begin{equation}
    \begin{split}
        \mathit{workload}(S_t) &= \sum_{q_i \in Q} \sum_{cf_j \in CF(P(q_i))} f_{i}(t) C_{ij} \delta_{ijt} \\ 
                      &+ \sum_{u_u \in U} \sum_{cf_n \in S(t)} f_{u}(t) C^{\prime}_{un} \delta_{nt} \label{WorkloadCost}
    \end{split}
\end{equation}
where $P(q_i)$ is a query plan group enumerated from $q_i$ and $CF(P(q_i))$ is a set of column families enumerated from $q_i$ used in query plan group $P(q_i)$\footnote{See Section~\ref{sec:proposal} for the detail of column family enumeration.}.
The first and the second terms on the right-hand side express query cost and update operation cost, respectively. 
The first term, $f_{i}(t)$ is the frequency of query $i$ at time step $t$ and $C_{ij}$ is the coefficient that represents the cost of query $i$ using column family $cf_j$.
$\delta_{ijt}$ is a binary decision variable which expresses whether query $q_i$ uses column family $cf_j$ at time step $t$.
Thus, the query cost is the summation of $f_{i}(t) C_{ij} \delta_{ijt}$ for all combinations of queries and column families.
The second term, $f_{u}(t)$ is the frequency of update operation $u$ at time step $t$ and $C^{\prime}_{un}$ is the coefficient that represents the cost of update operation $u$ for column family $cf_n$.
$\delta_{nt}$ is a binary decision variable which expresses whether column family $n$ exists in schema $S_t$ at time step $t$.
Thus, the update cost is the summation of $f_{u}(t) C^{\prime}_{un} \delta_{nt}$ for all combinations of update operation $u$ and column family $n$.

\subsection{Migration cost}
Remember that we choose a single optimized migration plan among multiple migration plan groups obtained from migration queries.
We define the database migration cost from old schema $S_{t}$ to new schema $S_{t + 1}$ using multiple migration queries as follows:
\begin{equation}
    \begin{split}
        migrate&(S_{t}, S_{t + 1}) =  \\
            &\sum_{cf_g \in S_{t + 1}}\sum_{cf_h \in CF(P(q_{o}^{\prime})), q_{o}^{\prime} \in M(cf_g)}  C^{E}_{h} \delta^{E}_{goht} \\
            & + \sum_{cf_g \in S_{t + 1}} C^{L}_{g} \delta^{L}_{g(t + 1)} \\ 
            & + \sum_{u_u \in U} \sum_{cf_g \in S_{t + 1}} C^{U}_{ug} \delta^{L}_{g(t + 1)}
        \label{migrateCost}
    \end{split}
\end{equation}
where $M(cf_g)$ is migration queries for target column family $cf_g \in S_{t + 1}$,
migration plan group $P(q_{o}^{\prime})$ and set of column families $CF(P(q_{o}^{\prime}))$
for migration query $q_{o}^{\prime}$ are similarly defined in the workload execution cost (Equation~\eqref{WorkloadCost}).
\par
The first term on the right-hand side expresses the cost of collecting records from old schema using migration plans.
$C^{E}_{h}$ is the coefficient that represents the cost of data collection from each column family $cf_h$.
$\delta^{E}_{goht}$ is a binary decision variable\footnote{$\delta^{E}_{goht}=0$ when $cf_g$ is not newly generated at time step $t$ for all $cf_h$.} that expresses whether migration query $q_{o}^{\prime}$ uses old column family $cf_h$ for generating new column family $cf_g$ at time step $t$.
Thus, the cost of collecting records from the old schema is the summation of $C^{E}_{h} \delta^{E}_{goht}$ for all combinations of column family $cf_g$, migration query $q_{o}^{\prime}$, and column family $cf_h$.

The second term expresses the cost of inserting the collected records into a new schema.
$C^{L}_{g}$ is the coefficient that represents the cost of inserting the collected records into new column family $cf_g$.
$\delta^{L}_{g(t + 1)}$ is a binary decision variable that expresses whether database migration to column family $cf_g$ is made between time step $t$ and $t + 1$. 
If $\delta^{L}_{g(t + 1)}=1$ then column family $cf_g$ does not exist at time step $t$ and exists at $t + 1$, so we introduce the following constraint~\eqref{Eq:CFMigration}: 
\begin{align} \label{Eq:CFMigration}
    \delta_{g(t + 1)} - \delta_{gt} \leq \delta^{L}_{g(t + 1)}
\end{align}

The third term expresses the cost of maintaining the new schema for ongoing update operations $U$ in workload.
$C^{U}_{ug}$ is the coefficient that represents the cost of update operation $u\in U$ for new column family $cf_g$ during the migration process.

\subsection{Constraints} \label{Section:Constraint}
We introduce three constraints for query plans, migration plans, and storage size
in order to choose column families required for optimized query/migration plans and avoid generating unused column families.

\subsubsection{Constraints for query plans}
Constraints for query plans ensure that 
for each query $q_i\in Q$ at every time step $t$, 
1) we choose a single optimized query plan $p_t$ among query plan group $P(q_i)$  (constraint~(\ref{const:ensureParentGeneral}, ~\ref{const:ensureOnePlanGeneral})), and 
2) all column families used in optimized query plan $p_t$ should exist in schema $S_{t}$ (constraint~\eqref{const:ensureExistenceGeneral}).
A decision variable is appropriately assigned to each $\delta_{ijt}$ and $\delta_{nt}$ in Equation~\eqref{WorkloadCost} using these constraints.

The first constraint~\eqref{const:ensureParentGeneral} ensures that if column family $cf_j$ is used in query plan $p_t$, any other column family $cf_l$ that precedes ($\prec$) $cf_j$ in the same query plan needs to be chosen.
\begin{equation}
    \forall cf_j, cf_l \in CF(P(q_i)). \:\: cf_l \prec_{p_t} cf_j \rightarrow \delta_{ilt} \geq \delta_{ijt}\label{const:ensureParentGeneral}
\end{equation}
where $\delta_{ijt}$ expresses whether query $q_i$ uses column family $cf_j$ at time step $t$ (introduced in Equation~\eqref{WorkloadCost}).

The second constraint~\eqref{const:ensureOnePlanGeneral} ensures that we produce a single unique query plan that joins all adjacent entities in a query graph.
\begin{equation}
    \begin{split}
        &\forall e, e^{\prime} \in \mathit{Entity}(q_i). \\ 
        &\quad \sum_{\substack{\{cf_j \in \mathit{CF}(P(q_i)) |\ e, e^{\prime} \in \mathit{Entity}(cf_j)\}}} \delta_{ijt} = 1 \label{const:ensureOnePlanGeneral}
    \end{split}
\end{equation}
where $\mathit{Entity}(q_i)$ is an entity set used in query $q_i$, $e$ and $e^{\prime}$ are entities that are adjacent in $q_i$'s query graph\footnote{The query graph expresses a partial schema relating to a given query extracted from the entity graph. Our current implementation is restricted to acyclic query graphs as in the implementation of NoSE~\cite{Mior2017}.}~\cite{Mior2017}, and
$\mathit{Entity}(cf_j)$ is a entity set from which column family $cf_j$ is generated.
The constraint~\eqref{const:ensureOnePlanGeneral} ensures that only single column family $cf_j$ is chosen from $CF(P(q_i))$ for partial columns of each adjacent entity pair $e, e^{\prime}$ used in $q_i$ at every time step $t$ (specified by $\sum{\delta_{ijt} = 1}$)\footnote{We permit generating a single column family for a leaf entity in the query graph\label{footnote_6}.}.

The third constraint~\eqref{const:ensureExistenceGeneral} ensures that if query plan $p_t$ is chosen as the optimized plan for $q_i$ at time step $t$, all column families ($cf_j$) used in $p_t$ should exist at the same time step.
\begin{equation}
    \forall cf_j \in CF(P(q_i)). \ \delta_{jt} \geq \delta_{ijt} \label{const:ensureExistenceGeneral}
\end{equation}
That is, if $\delta_{ijt}=1$ then $\delta_{jt}=1$ (remember that $\delta_{jt}$ was introduced in Equation~\eqref{WorkloadCost}).

As an example, we give the following constraints for $q_2$ at time step $t$ in Figure~\ref{TdQueryPlanExample}.
\begin{subequations}\label{Eq:OneTimeConstraint}
    \begin{align}
        \delta_{2,5,t} &\geq \delta_{2,3,t} \label{const:ensureParent}\\
        \delta_{2,4,t} + \delta_{2,5,t} &= 1, \delta_{2,3,t} + \delta_{2,4,t} = 1 \label{const:ensureOnePlan} \\
        \delta_{j,t}  &\geq \delta_{2,j,t} \;\;\; \forall j \in \{3, 4, 5\} \label{const:ensureExistence}
    \end{align}
\end{subequations}
where \eqref{const:ensureParent}, \eqref{const:ensureOnePlan}, and \eqref{const:ensureExistence} are instantiated from general constraints \eqref{const:ensureParentGeneral}, \eqref{const:ensureOnePlanGeneral}, and \eqref{const:ensureExistenceGeneral}, respectively.
Constraint~\eqref{const:ensureParent} ensures that $\mathit{CF}5$ is chosen whenever $\mathit{CF}3$ is used. 
The left-hand side of Constraint~\eqref{const:ensureOnePlan} specifies choosing either $\mathit{CF4}$ or $\mathit{CF}5$ for partial columns of an adjacent entity pair (\emph{Item} and \emph{User}), and the right-hand side specifies choosing either $\mathit{CF}3$ or $\mathit{CF}4$ for partial columns of a leaf entity (\emph{User}, see footnote\footref{footnote_6}).
Finally, constraint~\eqref{const:ensureExistence} ensures that $S_{t}$ contains $\mathit{CF}3, \mathit{CF}4, \mathit{CF}5$ if they are used in the optimized query plan of $q_2$.

\subsubsection{Constraints for migration plans}
\todo{The content is quite similar to the constraints for query plans. So, we may move this part to Appendix and give a short summary instead.}
Constraints for migration plans ensure that 
for each target column family $cf_g\in S_{t + 1}$ (specified by $\delta^{L}_{g(t + 1)} = 1$),
1) we choose a single migration plan based on the migration queries for the target column family $cf_g$ (constraints \eqref{const:ensureMigParentGeneral},\eqref{const:ensureMigOnePlanTreeGeneral},\eqref{const:ensureMigOnePlanGeneral}) and
2) all column families used in the chosen migration plan should exist in $S_{t}$ (constraint~\eqref{const:ensureMigExistenceGeneral}).
A decision variable is appropriately assigned to each $\delta^{E}_{goht}$ and $\delta^{L}_{g(t + 1)}$ in Equation~\eqref{migrateCost} using theses constraints.

Similarly to the constraint~\eqref{const:ensureParentGeneral} for query plans, 
the first constraint~\eqref{const:ensureMigParentGeneral} ensures that if the previous column family $cf_h$ is used in a migration plan, any other column family $cf_l$ that precedes ($\prec$) $cf_h$ in the same migration plan needs to be chosen.
\begin{equation}
    \begin{split}
        &\forall q^{\prime}_{o} \in M(cf_g), \; \forall cf_h, cf_l \in CF(P(q^{\prime}_{o})). \\ 
        &\qquad cf_l \prec cf_h \rightarrow \delta^{E}_{golt} \geq \delta^{E}_{goht}\label{const:ensureMigParentGeneral}
    \end{split}
\end{equation}
where $\delta^{E}_{golt}$ and $\delta^{E}_{goht}$ express whether old column family $cf_l$ and $cf_h$ exists in schema $S_t$, respectively.

We choose a single migration plan from migration queries for target column family $cf_g$ in two steps as follows.
In the first step, we choose single migration query $q^{\prime}_o$ from the migration queries $M(cf_g)$ (specified by $\sum\delta^{T}_{got}=1$) when target column family $cf_g$ is generated at time step $t+1$ (specified by $\delta^{L}_{g(t + 1)}=1$) using the second constraint~\eqref{const:ensureMigOnePlanTreeGeneral}.
\begin{equation}
    \sum_{q^{\prime}_{o} \in M(cf_g)} \delta^{T}_{got} = \delta^{L}_{g(t + 1)}
    \label{const:ensureMigOnePlanTreeGeneral}
\end{equation}

In the second step, we choose a single migration plan for the migration query chosen in the first step.
Similarly to the constraint~\eqref{const:ensureOnePlanGeneral} for query plans, the third constraint~\eqref{const:ensureMigOnePlanGeneral} ensures that only a single column family is chosen (specified by $\sum{\delta^{E}_{gomt} = \delta^{T}_{got}}$) for partial columns of adjacent entity pair $e, e^{\prime}$ in the migration query graph from each migration plan group transformed from $q_{o}^{\prime}$ at every time step $t$.
\begin{equation}
    \begin{split}
        &\forall e, e^{\prime} \in \mathit{Entity}(cf_g). \\
        &\qquad \sum_{\substack{\{cf_m \in \mathit{CF}(M(cf_g)) |\ e, e^{\prime} \in \mathit{Entity}(cf_m) \}}} \delta^{E}_{gomt} = \delta^{T}_{got} \label{const:ensureMigOnePlanGeneral}
    \end{split}
\end{equation}
where $\delta^{T}_{got}$ is a binary decision variable that expresses whether an optimized migration plan for the target column family $cf_g$ is chosen from migration plan group $P^{M}_o$ at time step $t$. 

Similarly to the constraint~\eqref{const:ensureExistenceGeneral}, the fourth constraint~\eqref{const:ensureMigExistenceGeneral} ensures that
if a migration plan is chosen as the optimized plan for generating $cf_g$ at time step $t$, all column families ($cf_h$) used in the optimized migration plan should exist at the same time step.
\begin{equation}
    \forall q^{\prime}_{o} \in M(cf_g), \forall cf_h \in \mathit{CF}(P(q^{\prime}_{o})). \:\: \delta_{ht} \geq \delta_{goh(t + 1)}^{E}\label{const:ensureMigExistenceGeneral}
\end{equation}

\begin{figure}[t]
    \begin{center}
        \includegraphics[scale = 0.97,bb = 0 0 212.88 101.04]{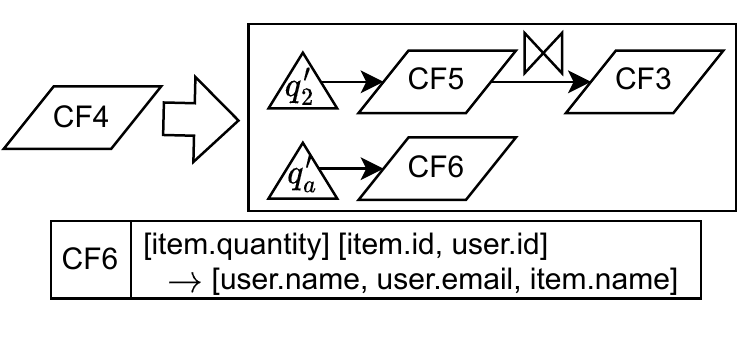}
        \caption{An example of enumerated migration plans for collecting data of $\mathit{CF}3$ in Fig.~\ref{TdQueryPlanExample}. In order to collect data for $\mathit{CF}3$, we generate migration query $q^{\prime}_1$, $q^{\prime}_2$ and $q^{\prime}_{a}$ and enumerate their migration plan groups using our proposed method (described in Section~\ref{Section:EnumerateMirgatePlan}). \todo{crop the figure}}
        \label{Fig:MigrationPlanExample}
    \end{center}
\end{figure}

Figure~\ref{Fig:MigrationPlanExample} depicts an example of a migration plan that generates a new column family $\mathit{CF}4$ in Figure~\ref{TdQueryPlanExample}.
The constraints to generate a single migration plan for $\mathit{CF}4$ are described as follows.
\begin{subequations}\label{Eq:PlanInMigPlanTree}
    \begin{align}
        \delta^{E}_{4,2,5,t} &\geq \delta^{E}_{4,2,3,t} \label{const:ensureMigParent} \\ 
        \delta^{T}_{4,a,t} &+ \delta^{T}_{4,2,t} = \delta^{L}_{4, t} \label{const:ensureMigOneTree} \\
        \delta^{E}_{4,2,5,t} = \delta^{T}_{4,2,t}&, \; \delta^{E}_{4,2,3,t} = \delta^{T}_{4,2,t}, \; \delta^{E}_{4,a,6,t} = \delta^{T}_{4,a,t} \label{const:ensureMigOnePlan} \\
        \delta_{3, t} \geq \delta^{E}_{4,2,3,t},\;& \delta_{5, t} \geq \delta^{E}_{4,2,5,t},\;
        \delta_{6, t} \geq \delta^{E}_{4,a,6,t} \label{const:ensureMigExistence}
    \end{align}
\end{subequations}
where \eqref{const:ensureMigParent}, \eqref{const:ensureMigOneTree}, 
 \eqref{const:ensureMigOnePlan}, \eqref{const:ensureMigExistence}, are instantiated from general constraints \eqref{const:ensureMigParentGeneral}, 
 \eqref{const:ensureMigOnePlanTreeGeneral}, \eqref{const:ensureMigOnePlanGeneral}, \eqref{const:ensureMigExistenceGeneral}, respectively.
Constraint~\eqref{const:ensureMigParent} ensures that the preceding $\mathit{CF}5$ is also chosen when $\mathit{CF}3$ is used in migration plan group 2. 
Constraint~\eqref{const:ensureMigOneTree} ensures that a single migration query is chosen among migration queries ($q^{\prime}_{a}$ and $q^{\prime}_{2}$).
Constraint~\eqref{const:ensureMigOnePlan} ensures that a single migration plan is chosen for $q^{\prime}_{a}$ and $q^{\prime}_{2}$, respectively.
Finally, constraint~\eqref{const:ensureMigExistence} ensures that $S_{t}$ contains all column families used in the optimized migration plan.

\subsubsection{Constraint for storage size} 

The storage size constraint~\eqref{baseObjective:spaceConst} ensures that the storage size of all column families should be smaller than $B$ at every time step $t$.
\begin{equation}
    \forall t \in [1, T]. \; \sum_j size(cf_j) \delta_{jt} \leq B \label{baseObjective:spaceConst}
\end{equation}
where $size(cf)$ is the storage size of column family $cf$. We ignore the size change of column families even when workload contains update operations, since the change in size is usually quite small compared to the whole database size.

%% file: parts/4_ProposedMethod.tex
\begin{figure*}[t]
  \begin{center}
    \includegraphics[scale = 0.48,bb = 0 0 1092 420.96]{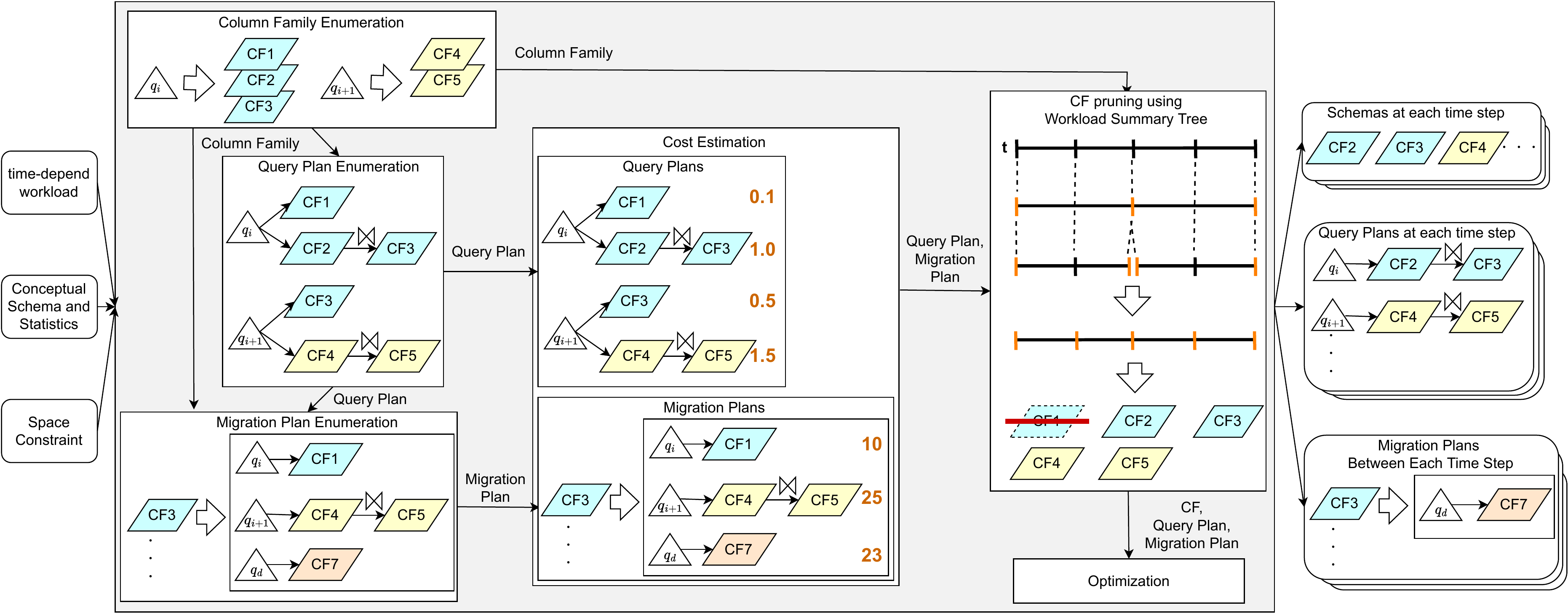}
    \caption{An overview of proposed method.\todo{simplify figure}}
    \label{SystemOverview}
  \end{center}
\end{figure*}
The optimization problem for time-series schema described in Section~\ref{Sec:Optimization} does not scale due to the large size of decision variables ($\delta_{jt}$ and $\delta^{E}_{goht}$).
% ($\delta_{jt}$, where $j$ is a column family and $t$ is a time step and $\delta^{E}_{goht}$ where $g, h$ are column family, $o$ is migration query, $t$ is a time step)
For example, the size of $\delta_{jt}$ is the product of the number of column families and the number of time steps.
This is caused by the fact that the number of schema candidates blows up depending on the number of time steps, and the number of migration plan candidates also blows up depending on the number of queries and schema candidates.
To overcome these obstacles, we propose a system for optimizing time-series schema by effectively reducing the number of schema (column family) candidates as well as the number of migration plan candidates.

%\memo{Proposal: 1st novelty}
First, we propose a novel column family pruning technique for time-dependent workload by introducing workload summary tree.
The workload summary tree approximates the ILP for the original workload using multi-level of workload summaries and the pruning technique effectively identifies uninteresting column families by leveraging the workload summary tree.
The novel idea is that we identify uninteresting column families that are never chosen as ILP answers at any nodes in the tree.
This approach is scalable because each summarized sub-workload at each node consists only of three time steps so we can largely reduce the size of decision variables for the ILP of each summarized sub-workload.

%\memo{Proposal: 2nd novelty}
Second, we propose an effective technique that reduces the number of migration plan candidates.
Notice that workload changes cause optimized query plan changes, which necessitate database migration. 
Our idea is that we can effectively reduce the number of migration plan candidates by restricting them according to how optimized query plans are changed instead of enumerating all possible candidates.

\subsection{System Overview}
Figure~\ref{SystemOverview} depicts an overview of our system.
Our system identifies an optimized time-series schema and outputs optimized query plans and migration plans using the following procedures: 
column family pruning (Section \ref{Section:PreOptimizationPruning}), 
column family and query plan enumeration (Section \ref{Section:EnumerateCFQueryPlan}), 
migration plan enumeration (Section \ref{Section:EnumerateMirgatePlan}), 
cost estimation (Section \ref{Section:Cost}), and 
optimization (Section \ref{sec:WholeOptimization}).

\subsection{Column family pruning using workload summary tree}
\label{Section:PreOptimizationPruning}
The objective function described in Section~\ref{Sec:Optimization} indicates that the decision variable size increases linearly to the time step size used in a time-dependent workload.
So, the optimization problem by ILP does not scale to a large number of time steps.
To tackle this issue, we propose a novel column family pruning technique for time-dependent workload by introducing novel hierarchical data structure, workload summary tree.
The workload summary tree approximates ILP of the original workload in multi-level of workload summaries and the pruning technique effectively identifies uninteresting column families by leveraging the multi-level of workload summaries.

\subsubsection{Workload summary tree}
We introduce workload summary tree in order to approximate ILP of the original workload.
We design the workload summary tree to have the following features, 1) every child node represents a sub-workload split in time-scale from the workload of its parent node, and 2) every parent node's workload is summarized from its children's sub-workload.
Therefore, the root node represents a highly summarized whole workload with small time steps and each leaf node represents an unsummarized sub-workload split. 
Each intermediate node represents a summarized sub-workload split.
In detail, we design the workload summary tree as a binary tree\footnote{To make the discussion simple, we assume the time step size in the workload is $2^n$. If not, we can add additional leaf nodes for the remainder time steps.}. 
Each node manages sub-workload with three time steps (minimum, median, maximum).
The workload managed at each parent node represents the summary of the workloads of its child nodes: the left child manages the left half of the parent workload (between the minimum and median time steps) and the right child manages the right half (between median and maximum time steps). 

\subsubsection{Column Family Pruning Algorithm}
By leveraging the novel workload summary tree, 
our pruning technique effectively identifies uninteresting column families using the multi-level of workload summaries: the upper level captures global aspects and the lower level captures more local aspect of the workload. 
Moreover, 
in order to take the global aspect from the upper level into account at the lower level, 
we propose a novel algorithm that recursively constructs an ILP for the summarized workload assigned to each node starting from the root and translates its answer as the ILP constraint of its child sub-workload.
In detail, the constraint enforces the ILP of child workloads to inherit optimized column families found at the parent workload: a child node solves local ILP so that the optimized column families at min/max time steps should be identical with the ones found at the same time steps (e.g. min and median for the left child node) at the parent workload. 
After computing the portion of the ILP for each node throughout the tree, we can identify uninteresting column families that are never chosen as ILP answers at any nodes in the tree.
This approach is efficient, because the time step size of the (sub-)workload at every node is limited to only three (minimum/median/maximum), which is significantly smaller than that of the original workload.

\begin{algorithm}[t]
\DontPrintSemicolon
\SetAlgoLined
\SetCommentSty{algocommfont}
\SetKwInOut{Input}{Input}\SetKwInOut{Output}{Output}
\SetKwProg{Fn}{Function}{ is}{end}
\Input{w (workload), const (ILP constraint)}
\Output{column families chosen at least by one node of workload summary tree}
\BlankLine
\Fn{get\_subtree\_cfs(w, const)}{
    \If{min\_ts + 1 is max\_ts}{
        \Return $\emptyset$\;
    }
    
    \tcp{get solution of current workload}
%    \State S \gets solve\_ILP(w, const)\; 
    $S \gets solve\_ILP(w, const)\;$
    
    \tcp{get used CFs in the solution}
%    \State C \gets S.cfs
   $C \gets S.cfs$

    \tcc{get middle time step of current workload}
%    \State mid\_ts \gets (w.min\_ts + w.max\_ts) / 2\;
    $mid\_ts \gets (w.min\_ts + w.max\_ts) / 2\;$

    \tcc{translate solution to the constraint}
%    \State const \gets const \bigcup translate\_to\_const(S)\;
    $const \gets const \bigcup translate\_to\_const(S)\;$
        
%    \State l\_child\_w \gets get\_child\_workload(w, w.min\_ts, mid\_ts)\;
    $l\_child\_w \gets get\_child\_workload(w, w.min\_ts, mid\_ts)\;$
    
    $C \gets C \bigcup get\_subtree\_cfs(l\_child\_w, const)\;$

%    \State r\_child\_w \gets get\_child\_workload(w, mid\_ts, w.max\_ts)\;
    $r\_child\_w \gets get\_child\_workload(w, mid\_ts, w.max\_ts)\;$
    
    $C \gets C \bigcup get\_subtree\_cfs(r\_child\_w, const)\;$
    
    \Return $C\;$
}
\caption{Obtaining interesting column families from subtree of workload summary tree}
\label{Algo:Presolve}
\end{algorithm}
The detailed algorithm is given in \ref{Algo:Presolve}.
$\mathit{solve\_ILP}()$ function (line 5) optimizes ILP for each node in the workload summary tree.
$\mathit{translate\_to\_const}()$ function (line 8) translates the ILP answer of a current node into the constraint for its child node in order to share the optimized column families found at $\mathit{min\_ts}$, $\mathit{mid\_ts}$, $\mathit{max\_ts}$ time steps. 
$\mathit{get\_child\_workload}()$ function (line 9, 11) constructs workload for a child node for given min/max time steps.
We recursively invoke $\mathit{get\_subtree\_cfs}()$ function (line 10, 12) by splitting the current workload into two child sub-workloads. 
After the recursion (line 3, 13), the identified interesting column families are returned. 
We remove column families that are not contained in these interesting column families from candidates of the optimization.

\subsection{Column Family And Query Plan Enumeration}\label{Section:EnumerateCFQueryPlan}

\subsubsection{Column Family Enumeration}\label{ssec:EnumerateCF}
The purpose of this step is to enumerate column family candidates used for the optimization problem of time-series schema (Section~\ref{Sec:Optimization}). 
Since arbitrary query plans can be chosen as optimized plans at any time step, we enumerate all possible column family candidates.
To this end, we take the same approach used in existing systems~\cite{Mior2017} for column family enumeration; We decompose each query and materialize the whole query or its sub-queries as column families. Thus, we can answer a query with a single column family (query efficient MV plan) or multiple column families by joining them (update efficient join plan). 
In detail, we employ a query graph, which is a sub-graph of the entity graph in a conceptual schema and expresses a partial schema referred from a given query.
We enumerate column family candidates by recursively decomposing query graphs; a query graph is decomposed at every node into two sub-queries that are materialized as column families.
In addition, in order to increase the utilization of column families for answering queries, we employ {\emph{relaxed queries}}~\cite{Mior2017} that are transformed from original queries by moving arbitrary attributes used in \texttt{WHERE}/\texttt{ORDER BY} clauses to \texttt{SELECT} clause\footnote{We keep at least a single equality predicate in \texttt{WHERE} clause in the same way as NoSE in order to construct a valid \texttt{Get} request for the column family.}.
The column families materialized from relaxed queries can be used to answer more queries, because they can answer queries with fewer conditions in \texttt{WHERE}/\texttt{ORDER BY} clauses.
However, the number of such column families increases exponentially with the number of query graph edges, because column families are materialized from sub-queries recursively decomposed at every edge.
Moreover, it increases relative to the factorial of the number of attributes used in \texttt{WHERE}/\texttt{ORDER BY} clauses, because column family variants are sensitive to attribute order. 
\par
To overcome such significant growth in the number of enumerated column families, we propose two pruning techniques.
First, we restrict the number of recursive decompositions of query graphs only to a single time for enumerating column families.
Thus, we only need to enumerate MV plans and two-CF join plans for all queries.
The enumerated query plans require at most a single join between column families.
The number of the original queries and decomposed sub-queries becomes $N_k = 1 + 2k$, which is linear to the number of edges ($k$) in a query graph.
Second, we reduce the variants of clustering keys in column families by following the features of extensible record stores: the prefix of clustering keys should contain attributes used in \texttt{GROUP BY}/\texttt{ORDER BY} clauses so that \texttt{GROUP BY}/\texttt{ORDER BY} operations can be executed on the server side.
Notice that we ignore the order of the remaining part of clustering keys if they do not appear in \texttt{WHERE}/\texttt{GROUP BY}/\texttt{ORDER BY} clauses.
Thus, we can reduce the number of column families by treating the remaining part of clustering keys to be order-insensitive.

\subsubsection{Query Plan Enumeration}

This step transforms each SQL query in the workload to query plans. % that are used in Section~\ref{ssec:QueryPlanAndMigrationPlan}.
We take the same approach proposed by Mior and Salem~\cite{Mior2017} as follows.
We enumerate query plans that join column family candidates enumerated in the column family enumeration step (Section~\ref{ssec:EnumerateCF}) from all queries. Here, each query plan corresponds to reconstructing the original query graph from its decomposed subqueries.
The enumerated query plans consist of three types of operations, 1) \texttt{Get}/\texttt{Put} operations for column families, 2) \texttt{ORDER BY}/\texttt{GROUP BY} at the server side, and 3) join/selection/\texttt{ORDER BY}/\texttt{GROUP BY} at the application side.

\subsection{Migration Plan Enumeration}\label{Section:EnumerateMirgatePlan}

The purpose of this step is to enumerate migration plan candidates used for the optimization problem of time-series schema (Section~\ref{Sec:Optimization}). 
That is, we enumerate all possible migration plans that migrate schema from $S_{t}$ to $S_{t+1}$ at any time step $t$. 
However, column families in $S_{t}$ are not decided before the schema optimization, so we enumerate migration plan candidates that generate each target column family enumerated in the column family enumeration (Section~\ref{Section:EnumerateCFQueryPlan}). 
We introduce two techniques for migration plan enumeration. 
The first one utilizes query plans as migration plans based on the fact that workload changes cause optimized query plan changes, which necessitate database migration. 
The second one enumerates additional migration plans to complement the first one.

\subsubsection{Migration plan enumeration by reusing query plans}
Optimized query plan changes require generating the target column families in $S_{t + 1}$ that are used by the optimized query plans at time step $t + 1$. 
We observe that if the optimized query plans at time step $t$ and $t+1$ are produced from the same query, 
they use column families (physical schema elements) produced from the same tables (conceptual schema elements), so the former plan outputs similar target column families in $S_{t + 1}$ as those of the latter plan.
Based on this observation, we can utilize query plans at time step $t$ as migration plans from $S_{t}$ to $S_{t + 1}$.
However, since optimized query plans are not decided before the schema optimization, we treat all enumerated query plans at time step $t$ as migration plan candidates in order to permit those query plans to become optimized query plans.

In detail, we identify source queries from workload queries $Q$ for each target column family in the enumerated column families (Section~\ref{Section:EnumerateCFQueryPlan}). 
A source query is a query whose query plans use the target column family. 
Then, we modify the source queries to migration queries by making the following modifications: 1) we simplify query plans by removing aggregation/\texttt{ORDER BY} operations that are not necessary for migration plans, and 2) we adjust the projections of the query plans to include the required columns of the target column family.
Finally, we enumerate query plans for the migration queries and treat them as migration plans.

Figure~\ref{Fig:MigrationPlanExample} depicts migration plan examples generated using $\mathit{CF}4$ in Figure~\ref{SchemaDesignExample} as the target column family.
Since $\mathit{CF}4$ is used in $q_2$ query plan group, we choose this as the source query for $\mathit{CF}4$. 
Then, we modify $q_2$ to migration query $q^{\prime}_2$ as follows:
\begin{verbatim}
SELECT item.id, user.id, user.name, user.email
FROM user.item WHERE item.quantity = ?
\end{verbatim}
$q^{\prime}_2$ shares the same \texttt{FROM}/\texttt{WHERE} clauses as $q_2$'s but with a different \texttt{SELECT} clause which contains the necessary columns of $\mathit{CF}4$.
Finally, we generate migration plans using $q^{\prime}_2$ and $\mathit{CF}3$ and $\mathit{CF}5$ (excluding the target column family, $\mathit{CF}4$).
We obtain $p_{2,2}$ as a migration plan which uses $\mathit{CF}3$ and $\mathit{CF}5$.

\subsubsection{Complementary migration plan enumeration}
The above enumeration technique may not enumerate appropriate migration plans when the number of query plans in each query is small. 
To complement this, the second technique enumerates additional migration plans using a simple migration query, which specifies the partition key and columns of the target column family in WHERE clause and SELECT clause, respectively. 
In Fig.~\ref{Fig:MigrationPlanExample}, $q^{\prime}_a$ is simple migration query of $\mathit{CF}4$ and the second technique enumerates its migration plan uses $\mathit{CF}6$.

\subsection{Cost Estimation} \label{Section:Cost}
In order to optimize the objective of Equation~\ref{baseObjective:Objective}, we need to estimate the coefficients used in the workload execution cost (Equation~\ref{WorkloadCost}) and migration cost (Equation~\ref{migrateCost}).
\par
Remember that $C_{ij}$ in Equation~\ref{WorkloadCost} is defined as the coefficient that represents the cost of query $i$ processing using column family $cf_j$. We estimate it using linear regression with the function $T(n, w, s)$\footnote{We choose the simplest approach of using linear regression. We can utilize more recent techniques using deep learning for achieving higher accuracy.} where $n$ is the number of \texttt{Get} operations, $w$ is the query-cardinality (the expected number of records in the result), and $s$ is the record size of column family $j$. 
$n$ is computed as the sum of \texttt{Get} operations in the query plan tree: a single \texttt{Get} operation is required for the first step in the tree, and we use the query-cardinality of this node as the required number of \texttt{Get} operations for later steps, because we need to invoke a \texttt{Get} operation for each record returned from this node.
$w$ is computed using the cardinality of attributes in equality conditions and using the number of records of each entity in the conceptual schema.
When query $i$ uses a \texttt{GROUP BY} clause, $w$ is computed using the number of expected groups by pushing down the \texttt{GROUP BY} clause at the server side.
Since linear regression function $T(n, w, s)$ depends on the instance of extensible record stores and its computing environment, we train $T(n, w, s)$ using performance profiles as the training datasets collected by changing queries (attributes used in \texttt{SELECT}/\texttt{WHERE} clauses) and the number of records in column families.
Next, $C^{\prime}_{un}$ in Equation~\ref{WorkloadCost} is defined as the coefficient that represents the cost of update operation $u$ for column family $cf_n$.
We estimate it using linear regression function $T^{\prime}(w)$ where $w$ is the number of expected updated records. 
Similar to the above $C_{ij}$ estimation, we train $T^{\prime}(w)$ using performance profiles.

As for the migration cost, we need to estimate the coefficients $C^{E}, C^{L}, C^{U}$ used in Equation~\ref{migrateCost}.
First, $C^{E}_{h}$ is the coefficient that represents the cost of data collection from each column family $cf_h$, which depends on the size of $cf_h$.
%マイグレーションプランの実行時には，プラン内で使用する column family の全レコードをそれぞれ収集するため\footnote{マイグレーションプランがジョインプランである場合にはクライアントにおいてジョイン処理を行う．このジョイン処理はデータベースの処理を圧迫しないため，ジョイン処理のコストは考慮しない．}，$C^{E}_{gh}$ はレコードを収集する対象の column family $h$ のサイズに依存する．
So, we estimate $C^{E}_{h}$ using linear regression with the function $T^{\prime\prime}(s_h)$ where $s_h$ is the size of $cf_h$.
Second, $C^{L}_{g}$ is the coefficient that represents the cost of inserting the collected records into new column family $cf_g$, which depends on the size of $s_g$.
So, we estimate $C^{L}_{g}$ using linear regression with the function $T^{\prime\prime\prime}(s_g)$ where $s_g$ is the size of $cf_g$.
Finally, $C^{U}_{ug}$ is the coefficient that represents the cost of the update operation $u\in U$ for new column family $cf_g$ during the migration process. 
We estimate it using Equation~\eqref{Eq:UpdateConstructingCF}:
\begin{equation}\label{Eq:UpdateConstructingCF}
  C^{U}_{ug} = f_{u(t - 1)} C^{\prime}_{ug} \frac{C^{L}_{g}}{(interval)}
\end{equation}
where $C^{L}_{g} / (interval)$ is the time ratio of column family $g$ generation to the interval between time steps, and 
$f_{u}(t-1)$ is the frequency of update operation $u$ at time step $t-1$.

Similar to the coefficient estimation for workload cost, we train the above regression models using performance profiles from collecting and inserting records as the training datasets: the profiles are collected by changing the number of records in column families.

\subsection{Optimization}\label{sec:WholeOptimization}
Finally, we obtain the optimal design for the time-series schema by minimizing the total cost of the time-dependent workload execution and database migrations for enumerated column families, query plans, and migration plans. 
In addition, we additionally minimize the number of column families and their storage size with three steps as follows.

First, we identify the optimal design for the time-series schema using Equation~\eqref{baseObjective:Objective} and keep its minimum cost. 
Second, we additionally minimize the number of column families using \eqref{eq:minCFNum} while keeping the cost of Equation~\eqref{baseObjective:Objective} as the minimum cost.
\begin{equation}\label{eq:minCFNum}
    \min \sum_{t = 1}^{T} \sum_{j} \delta_{jt} 
\end{equation}
Finally, we also minimize the size of column families using \eqref{eq:minStorage} while keeping the number of column families as the minimum.
\begin{equation}\label{eq:minStorage}
    \min \sum_{t = 1}^{T} \sum_{j} s_{j}\delta_{jt} 
\end{equation}
This approach is especially effective when the workload does not have update operations, because the approach ensures the minimality of the number of column families and their size.

\begin{comment}
\subsection{Limitation}\label{Section:limitation}
Our system does not fully support the specification of SQL languages.
First, our system does not handle queries without equal conditions.
Second, it does not support several predicates, such as LIKE and BETWEEN.
Third, it does not support nested queries.
We have detours for the second and third limitations; rewriting some predicates to equal and range conditions, and decomposing a nested query into multiple flat queries. 
We use these detours in our experiments when workloads face these limitations.
We note that NoSE also has the same limitations. Furthermore, NoSE cannot support aggregation functions such as SUM and AVERAGE.
Thus, our system supports richer queries than NoSE.
\end{comment}

%% file: parts/5_Experiments.tex
We make experiments in order to answer the following questions. 
Q1: we evaluate the effectiveness of our proposal; how largely our proposal improves the workload latency compared to NoSE, which optimizes the schema design by approximating a time-depended workload with an average static workload (Section \ref{sec:evalLatency}).  
Q2: we evaluate the effectiveness of workload summary tree, that is how largely the optimization execution time is reduced using workload summary tree (Section \ref{sec:evalOptRunningTime}). 
Q3: 
%この章では，提案手法の性能を評価するために3つの問いを設けて，それぞれ評価する．
%1つ目の問いとして，提案手法がスキーマのマイグレーションを実行することで，NoSE を用いたスキーマよりも応答時間を低減しているかを確認する (\ref{sec:evalLatency}節)．
% 2つ目の問いとして，抽象化による候補削減手法によって，最適化の実行時間を低減していることを確認する (\ref{sec:evalOptRunningTime}節)．
そして，3つ目の問いとして，提案手法の効率的な CF，クエリプラン列挙手法を用いることで，NoSE に比べてスキーマ最適化の実行時間を低減しているかを確認する (\ref{sec:evalCFEnumerationTime}節)．

All experiments were performed on a single server with two Intel(R) Xeon(R) Gold 6130 CPU (2.10GHz processors, 64 logical cores) and 1.5TB main memory using Docker vXXXXXX with Ubuntu XX.XX.
We use Apache Cassandra (version 3.11.8) as an extensible record store. 
% 本稿の性能評価では，extensible record store の一つである Apache Cassandra（バージョンは 3.11.8）を利用し， CPU に 2.10GHz の Intel(R) Xeon(R) Gold 6130 CPU を2つ搭載し，1.5TB のメモリを持つサーバを使用した\todo{ストレージも確認．NoSE くらいの粒度で説明したい}．
We make experiments using a Docker container of five nodes of Apache Cassandra on the single server
and evaluate workload latency.  
1台のサーバ上で5つのノードを持つ Apache Cassandra を Docker コンテナで構築し，クライアントとマイグレータも同一のサーバ上で動作させて応答時間を計測した．
本稿で提案するコストモデルは Cache の影響を考慮していないため，Cassandra でのキャッシュは全て無効化した．
また，ILP のソルバはバージョン 10.0.1 の Gurobi~\footnote{https://www.gurobi.com/} を用いた．

\subsection{Experimental Setting}\label{sec:setting}

\begin{figure}[t]
    \begin{minipage}[b]{0.5\linewidth}
        \includegraphics[scale = 0.31, bb = 0 0 726 275]{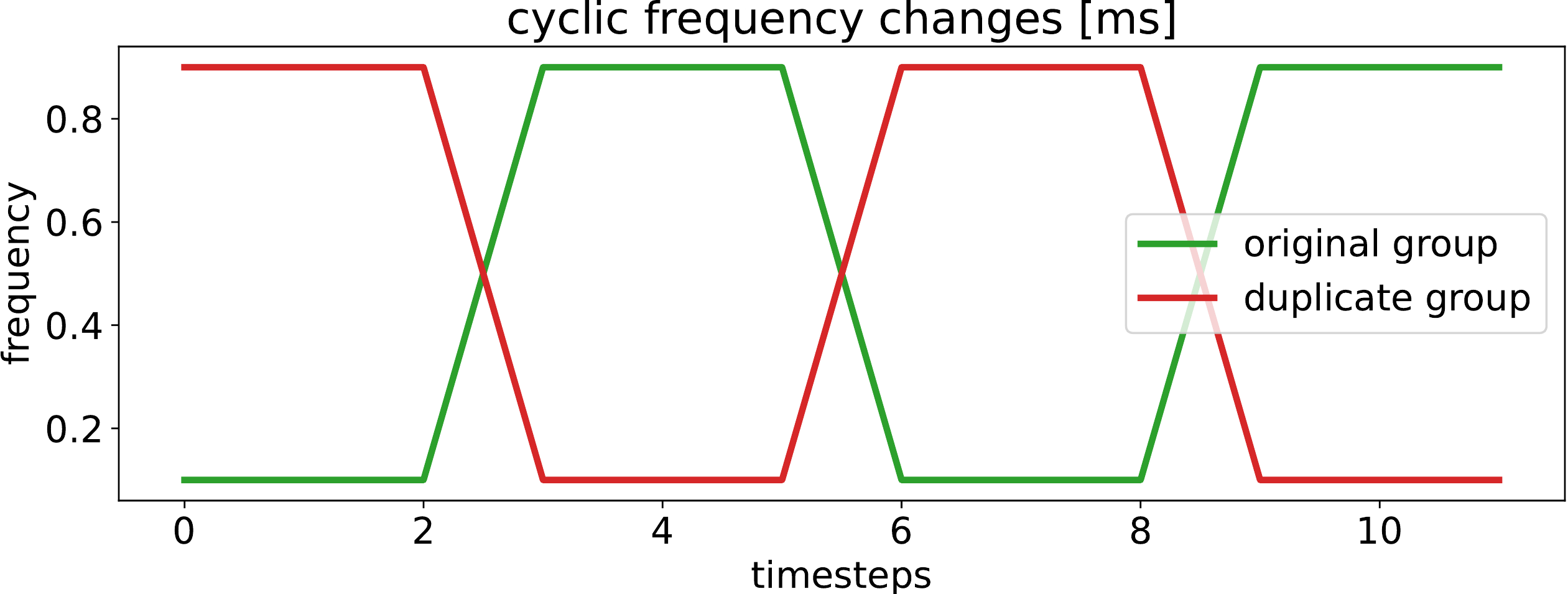}
    \end{minipage} \\
    \begin{minipage}[b]{0.5\linewidth}
        \includegraphics[scale = 0.31, bb = 0 0 726 275]{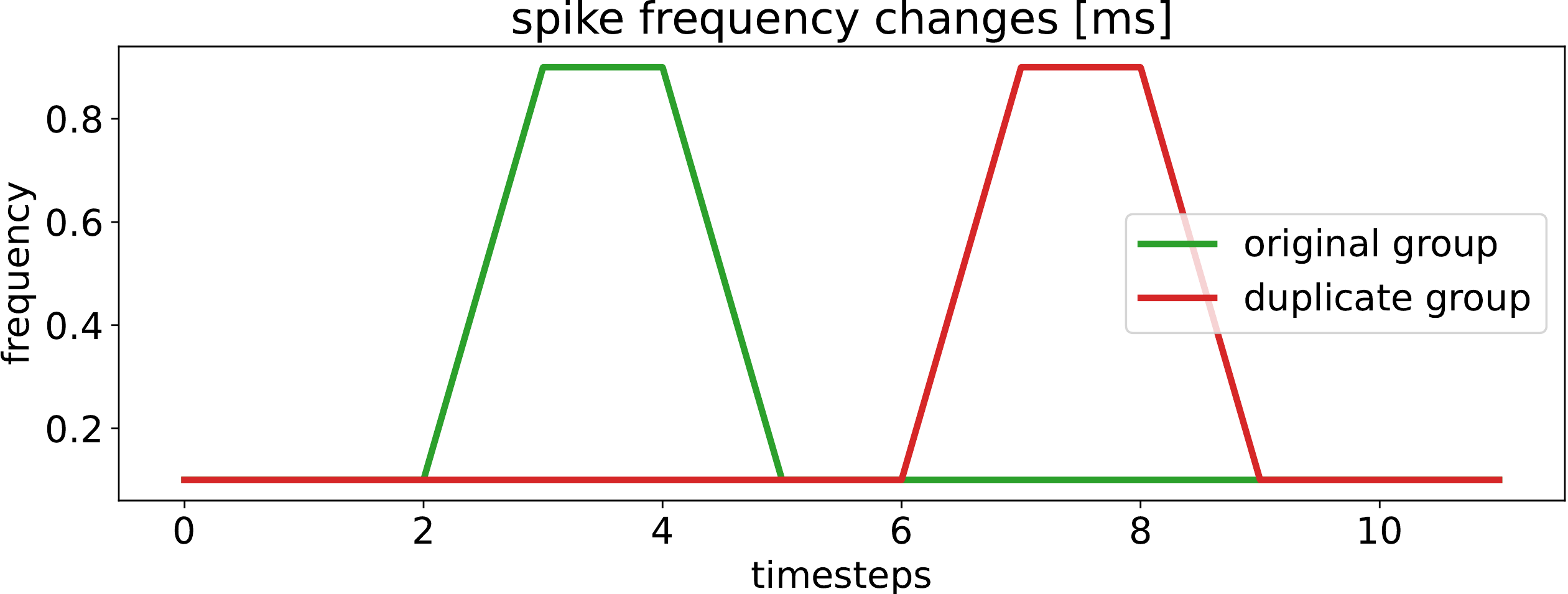}
    \end{minipage} \\
    \begin{minipage}[b]{0.5\linewidth}
        \includegraphics[scale = 0.31, bb = 0 0  726 275]{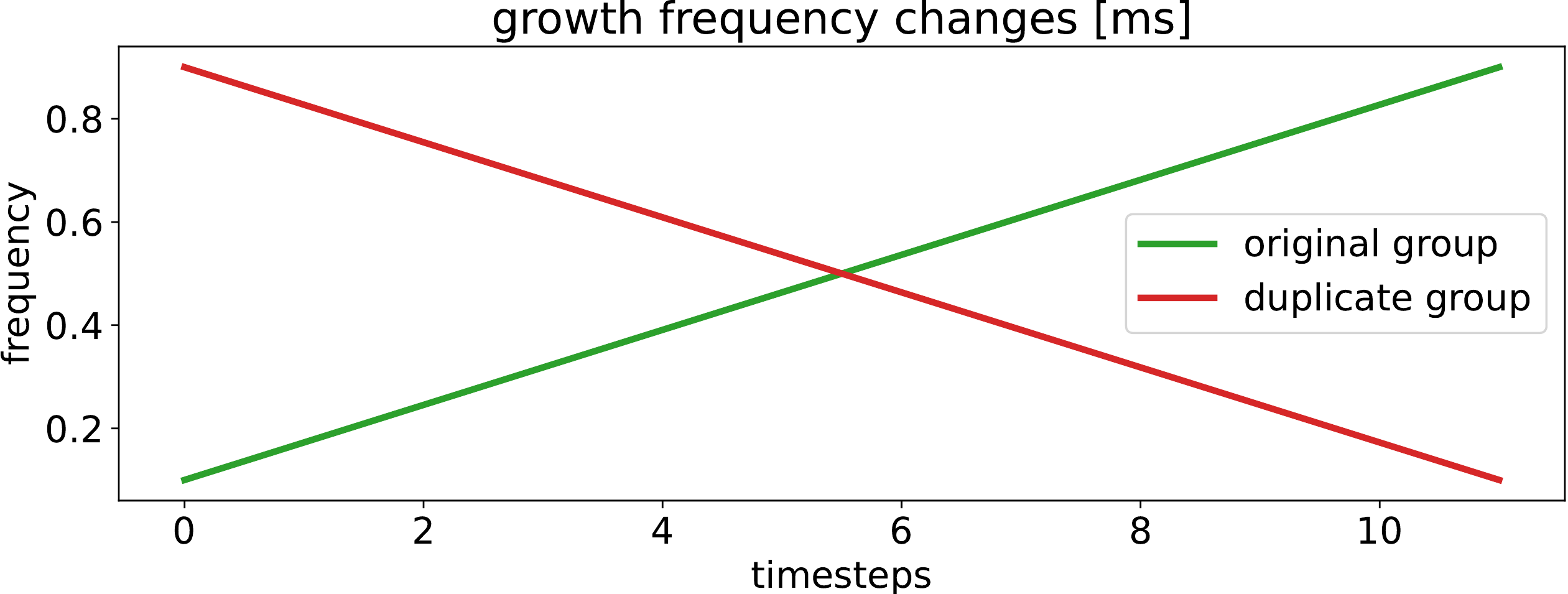}
    \end{minipage}
    \caption{各時刻変化するワークロードのクエリグループの実行頻度の変化．}
    \label{eval:FreqChange}
\end{figure}

{\bf Workload: } 
Since there are not open real-world time-dependent workloads, we generate two synthetic time-dependent workloads by modifying two static benchmarks, TPC-H~\footnote{http://www.tpc.org/tpch/} and RUBiS~\cite{Cecchet2002};
We change the query frequencies by following the typical time-depended settings summarized in prior study~\cite{Ma2018}.
We use three types of ratio changes:
\begin{itemize}
    \item Periodical: The query ratios periodically increase and decrease. 各時刻のユーザ行動の傾向によって問い合わせが変化するサービスでは，一日毎に周期的に問い合わせの頻度が変化する場合がある．このような頻度変化において，企業が経験則から頻度変化を予測しスキーマをマイグレーションする場合を想定する．
    \item Spike:時刻変化においてスパイクが発生する場合を模した頻度変化である．
この頻度変化は，サービスにおいて，情報公開等による急激なアクセス増加を企業が予測し，スキーマをマイグレーションする場合を想定する．
    \item Linear: ある処理の実行頻度が単調に変化する場合を模した頻度変化である．
これは，期限のある手続きに対して期限前に近づくほど単調に問い合わせ回数が増加するような場合を想定している．
\end{itemize}
各ワークロードのそれぞれのグループの実行頻度の変化を図~\ref{eval:FreqChange} に示す．
%本実験では，意思決定支援システムを模したベンチマークである TPC-H~\footnote{http://www.tpc.org/tpch/} に時刻変化を加えたワークロードを用いる．
%\footnote{NoSQL データベースのベンチマークとして YCSB~\cite{Cooper2010}やRUBiS~\cite{Cecchet2002}が使用されている．しかし，YCSB は限定的なスキーマのみを用いておりスキーマ設計の性能評価には適さない．RUBiS はシンプルなクエリのみを持つため，スキーマの選択肢が十分に無く，マイグレーションの性能評価には適していない．}
% ここで，提案手法の対応するクエリ言語は NoSE のクエリ言語に集約処理等の機能を追加したものであり，SQL に比べて限定的な機能のみを持つ．
% そのため，主に以下の3つの変更を加えることで，TPC-H のクエリを処理可能なクエリへ変換した．
% 1つ目の変更として，NoSE と同様に，等号条件を持たないクエリに対応していないため一部のクエリを削除した．
% 2つ目の変更として，提案手法のクエリ言語は LIKE や BETWEEN などをサポートしていないため，それぞれ等号条件や範囲条件に変更した．
% 3つ目の変更として，入れ子クエリも個別のクエリへ分解して，等号条件を持つクエリのみワークロードに個別のクエリとして追加した
TPC-H のレコードは，スケールファクタが 1 のレコードを作成して用いた．
% \par
% TPC-H は静的なベンチマークであり時刻変化しないため，本実験では3種類の頻度変化を加えてベンチマークに用いた．
% これらの頻度変化は，Ma ら~\cite{Ma2018} の用いたワークロードを模したものである．
%ただし，Ma らの用いた実データは公開されていないため，簡単化した頻度変化を生成して用いた．
% 1つ目の頻度変化は，周期的に実行頻度が変化する場合を模した頻度変化である．

% 2つ目の頻度変化は，
% 3つ目の頻度変化は，
% また，本稿で対象とする時刻変化は頻度変化のみを対象とするため，ワークロードのクエリに含まれるクエリは一定とした．

\par

時刻変化するワークロードにおいて，全てのクエリの実行頻度が同様に変化した場合，マイグレーションの必要性は小さく，性能評価に適さない．
そこで，オリジナルの TPC-H のクエリを持つグループと，複製した TPC-H のクエリを持つグループの2グループを持つワークロードを作成した．
そして，これらのグループに異なる周期変化の周期，ピークの時刻，単調変化の増減を割り当てて，各時刻において重要となる問い合わせが変化するワークロードを作成した．
各ワークロードのそれぞれのグループの実行頻度の変化を図~\ref{eval:FreqChange} に示す．

{\bf Comparison: } 
提案手法の比較手法として，5つの手法を実装した．
\begin{itemize}
\item \textbf{平均実行頻度に対する最適化}: 
各時刻の実行頻度の平均値を実行頻度として用いる最適化である．
\item \textbf{最小時刻実行頻度に対する最適化}: ワークロードの最小時刻の実行頻度を用いる最適化である．
\item \textbf{最大時刻実行頻度に対する最適化}: ワークロードの最大時刻の実行頻度を用いる最適化である．
\item \textbf{候補削減無しの提案手法}: 提案手法の抽象化による候補削減を無効化し，全ての候補について一括で最適化する．
この手法では最適化の実行時間は増加するが，より精度の高いスキーマを期待できる．
\item \textbf{理想的な最適化}:マイグレーションを考慮せず各時刻のスキーマを独立に最適化することで，各時刻のワークロードに最も適したスキーマを推薦する．
しかし，マイグレーションを想定していないため更新処理の結果等を引き継ぐことができない．
さらに，マイグレーションが出来ないため，各時刻のスキーマを他の時刻でも保持しなければならず，ストレージを圧迫する．
したがって，実際にデータベースを運用する際にこの手法を用いることは困難であるが，提案手法の性能の上限値の目安として用いる．
\end{itemize}
平均，最小，最大時刻実行頻度に対する最適化は全ての時刻において同じスキーマを使用する静的な最適化手法であり，NoSEに相当する．
NoSEを直接利用する場合，TPC-H の複雑なクエリに対してスキーマ決定が困難であったため，CF列挙・クエリプラン列挙・コストモデル・目的関数は提案手法と同様の手法を使用した．

{\bf Metrics and parameters: } 
本稿の実験では各クエリをその実行頻度によらず，全ての時刻で同一の回数実行する．
そして，本来の実行頻度に応じて重みを付けてその平均値を求めることで，各時刻の実行頻度による応答時間の変化を考慮する．

本実験で用いるワークロードは，更新処理を持たない TPC-H をベースとしているためストレージサイズの制約を設ける．
ストレージサイズの制約を設けることで，全てのクエリに対して MV プランを推薦できず，最適なスキーマ設計が困難となる場合において提案手法の性能を確認する．
ストレージサイズの制約の値を決定するために，制約を設けず平均実行頻度に対する最適化を実行し，そのストレージサイズの推定値を取得する．
そして，その推定値から一定の割合削減したサイズをストレージサイズの推定値の制約値として用いる\footnote{ストレージサイズの制約は，スキーマが正規化されるという点において，更新処理を考慮すると同様の影響を持つ．}．

\subsection{ワークロードの応答時間の評価}\label{sec:evalLatency}
本節では，時刻変化するワークロードに対して提案手法が適切にマイグレーションを実行して，応答時間を低減するかを確認する．

\subsubsection{実行頻度が周期的に変化するワークロードでの実験}\label{sec:periodical_latency}
\begin{figure}[t]
    \includegraphics[scale = 0.62, bb= 0 0 409 256]{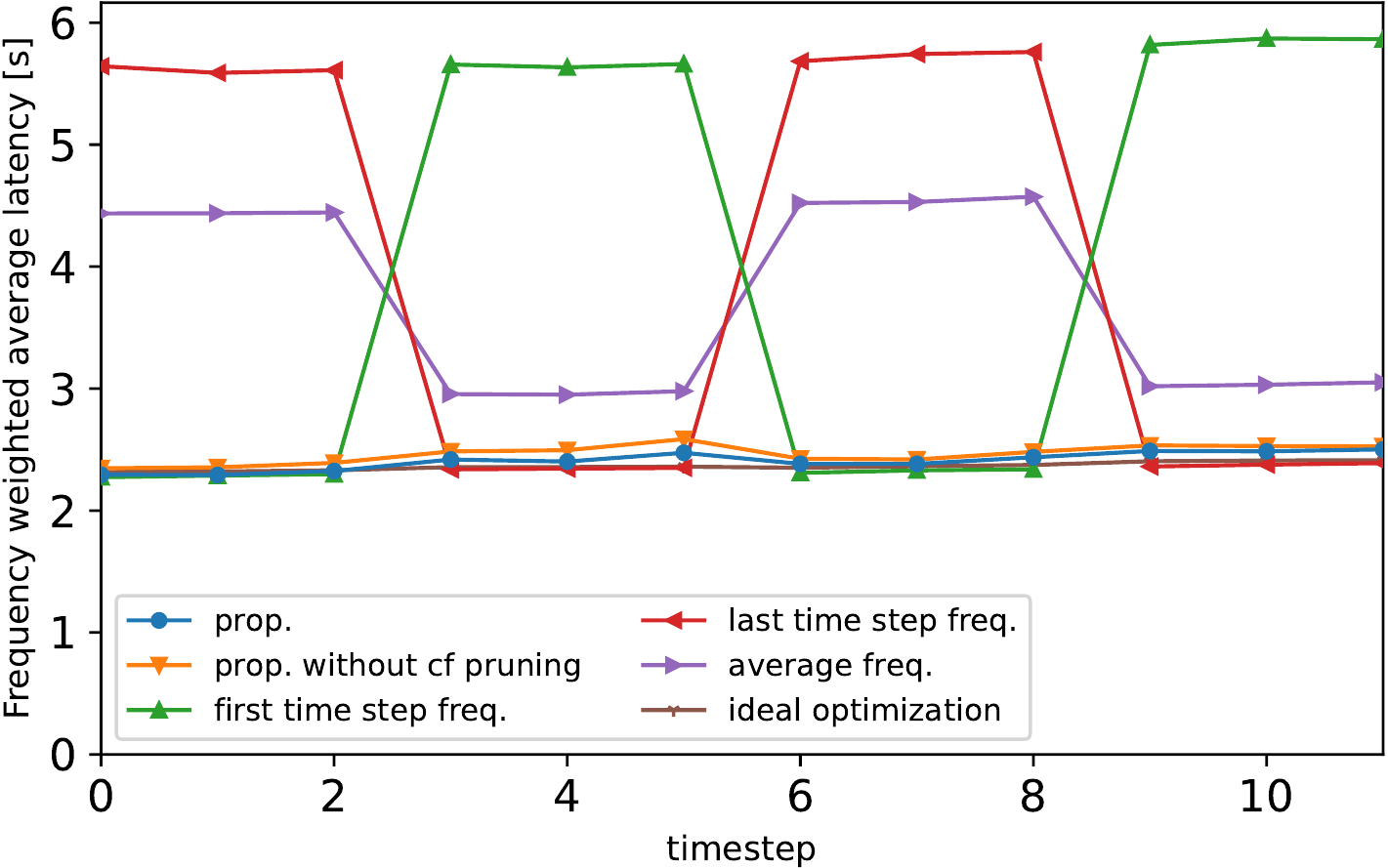}
    \caption{実行頻度が周期的に変化するワークロードでの応答時間の推移．横軸が各時刻，縦軸が応答時間の実行頻度による加重平均．静的な最適化手法では実行頻度の変化に追従できず，応答時間の加重平均値が増加しているが，提案手法は低い応答時間の加重平均値を維持している．}
    \label{eval:cyclic_latency}
\end{figure}

図~\ref{eval:FreqChange}の実行頻度が Periodical に変化するワークロードに対して，各手法の推薦するスキーマの応答時間を計測した結果を図~\ref{eval:cyclic_latency} に示す．
ストレージ容量の制約値は，ストレージ制約の無い平均実行頻度の最適化のストレージ値の 80\% のサイズを用いた．
提案手法はマイグレーションを活用することで，ワークロードの実行頻度が Periodical に変化する場合でも低い応答時間を維持することを確認した．
また，時刻 0,1,2,6,7,8 では理想的な最適化と同程度の応答時間を達成した．
提案手法の実行頻度による加重応答時間平均は，最小時刻実行頻度に対する最適化に比べて40.2\%，最大時刻実行頻度に対する最適化に比べて 40.0\%, 平均実行頻度に対する最適化に比べて 35.7\%削減していることを確認した．
また，理想的な最適化に比べると実行頻度による加重応答時間平均は22.9\% 増加した．
候補削減無しの提案手法は提案手法に比べてより多くの最適化候補を用いて最適化する．
そのため，提案手法よりも候補削減無しの提案手法の方がより応答時間を低減すると想定される．
しかし実験では，候補削減無しの提案手法に比べ，実行頻度による加重応答時間平均を2.3\%低減した．
これは，提案手法は候補削減によって，限定的な最適化候補のみを用いるため，提案手法の応答時間の方が候補削減無しの提案手法よりも応答時間が大きくなるという想定に反する．
原因としては，提案手法のシンプルなコストモデルでは，各属性のカーディナリティが十分な精度で推定できず応答時間に差異が生じている可能性が考えられる．

各比較手法について，最小時刻実行頻度に対する最適化や最大時刻実行頻度に対する最適化は，それぞれ時刻0や時刻11で応答時間を低減している．
一方で，最適化対象とした時刻と実行頻度が異なる時刻においては応答時間が大幅に増加している．
平均実行頻度に対する最適化では，最小時刻実行頻度に対する最適化と最大時刻実行頻度に対する最適化に比べて低い応答時間を維持しているが，実行頻度の変化に応じて応答時間が増加している時刻がある．

\todo{最大時刻実行頻度に対する最適化が時刻11に理想的な最適化に負けている理由を考察する．表\ref{eval:periodical_mig_plans} に示したマイグレーションプランについての考察に1段落分割く}

\begin{table}[]
\begin{tabular}{l|l|l}
time step & cf migration plan type & query \\ \hline
                                     & MV Plan $\rightarrow$ Join Plan  & Q3-dup, Q5-dup, Q7-dup, Q9-dup \\ \cline{2-3} 
\multirow{-2}{*}{2 $\rightarrow$ 3}  & Join Plan $\rightarrow$  MV Plan & Q3, Q5, Q7, Q9                 \\ \hline
                                     & MV Plan $\rightarrow$ Join Plan  & Q3, Q5, Q7, Q9                 \\ \cline{2-3} 
\multirow{-2}{*}{5 $\rightarrow$ 6}  & Join Plan $\rightarrow$ MV Plan  & Q3-dup, Q5-dup, Q7-dup, Q9-dup \\ \hline
                                     & MV Plan $\rightarrow$ Join Plan  & Q3-dup, Q5-dup, Q7-dup, Q9-dup \\ \cline{2-3} 
\multirow{-2}{*}{8 $\rightarrow$ 9}  & Join Plan $\rightarrow$ MV Plan  & Q3, Q5, Q7, Q9                
\end{tabular}
\caption{容量制約80\%下で Periodical に実行頻度が変化するワークロードを最適化した際に提案手法が推薦した cf migration plan}
\label{eval:periodical_mig_plans}
\end{table}

\begin{table}[]
\begin{tabular}{l|l|l}
time step & cf migration plan type & query \\ \hline
                                     & MV Plan $\rightarrow$ Join Plan  & Q3-dup, Q5-dup, Q7-dup, Q9-dup \\ \cline{2-3} 
\multirow{-2}{*}{5 $\rightarrow$ 6}  & Join Plan $\rightarrow$ MV Plan  & Q3, Q5, Q7, Q9 \\ \hline
\end{tabular}
\caption{容量制約80\%下で Linear に実行頻度が変化するワークロードを最適化した際に提案手法が推薦した cf migration plan}
\label{eval:Linear_mig_plans}
\end{table}

\begin{figure}[t]
    \includegraphics[scale = 0.62, bb = 0 0 409 256]{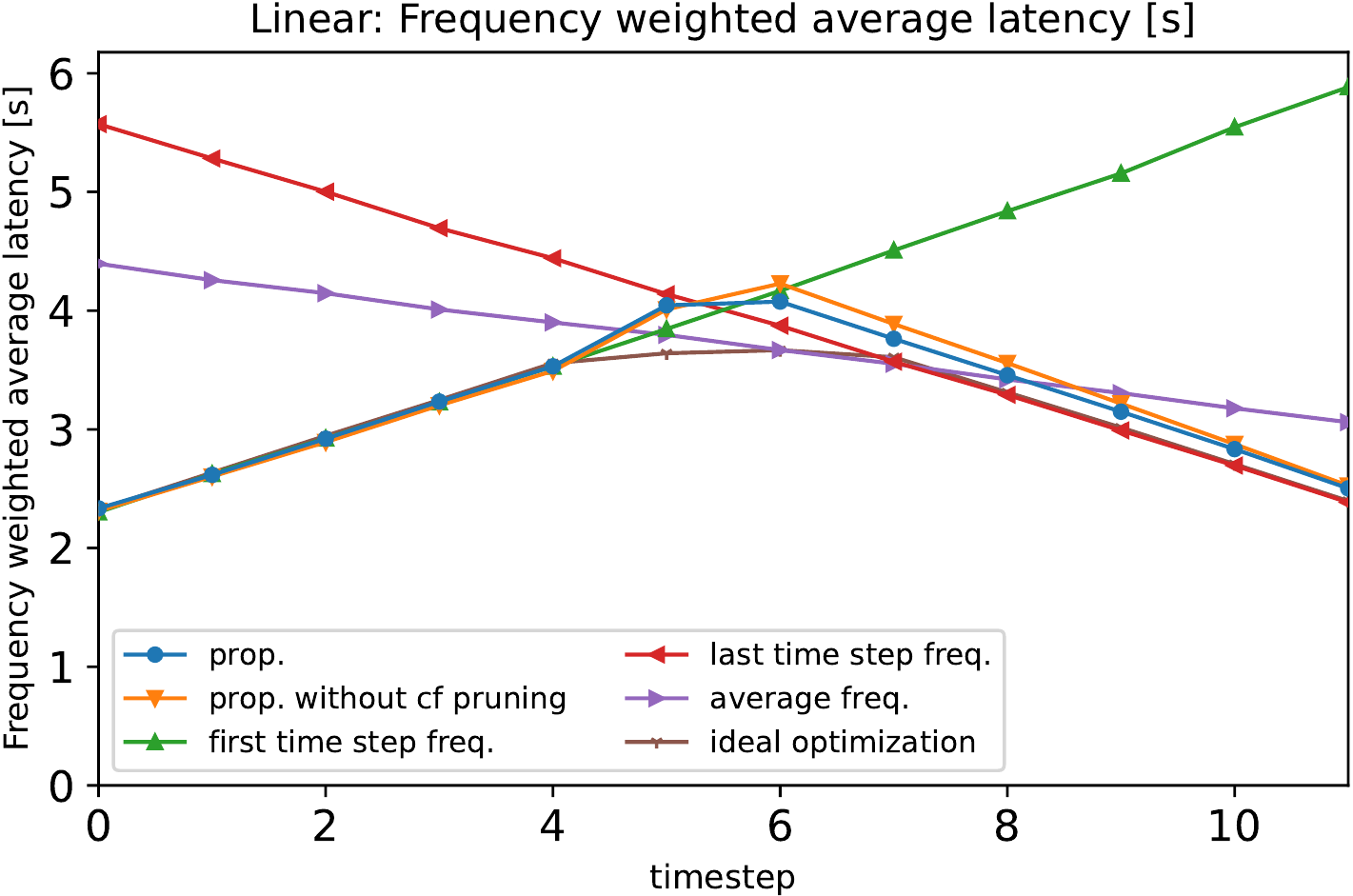}
    \caption{Linear, 80per, }
    \label{eval:monotonic_latency}
\end{figure}
\begin{figure}[t]
    \includegraphics[scale = 0.62, bb = 0 0 409 256]{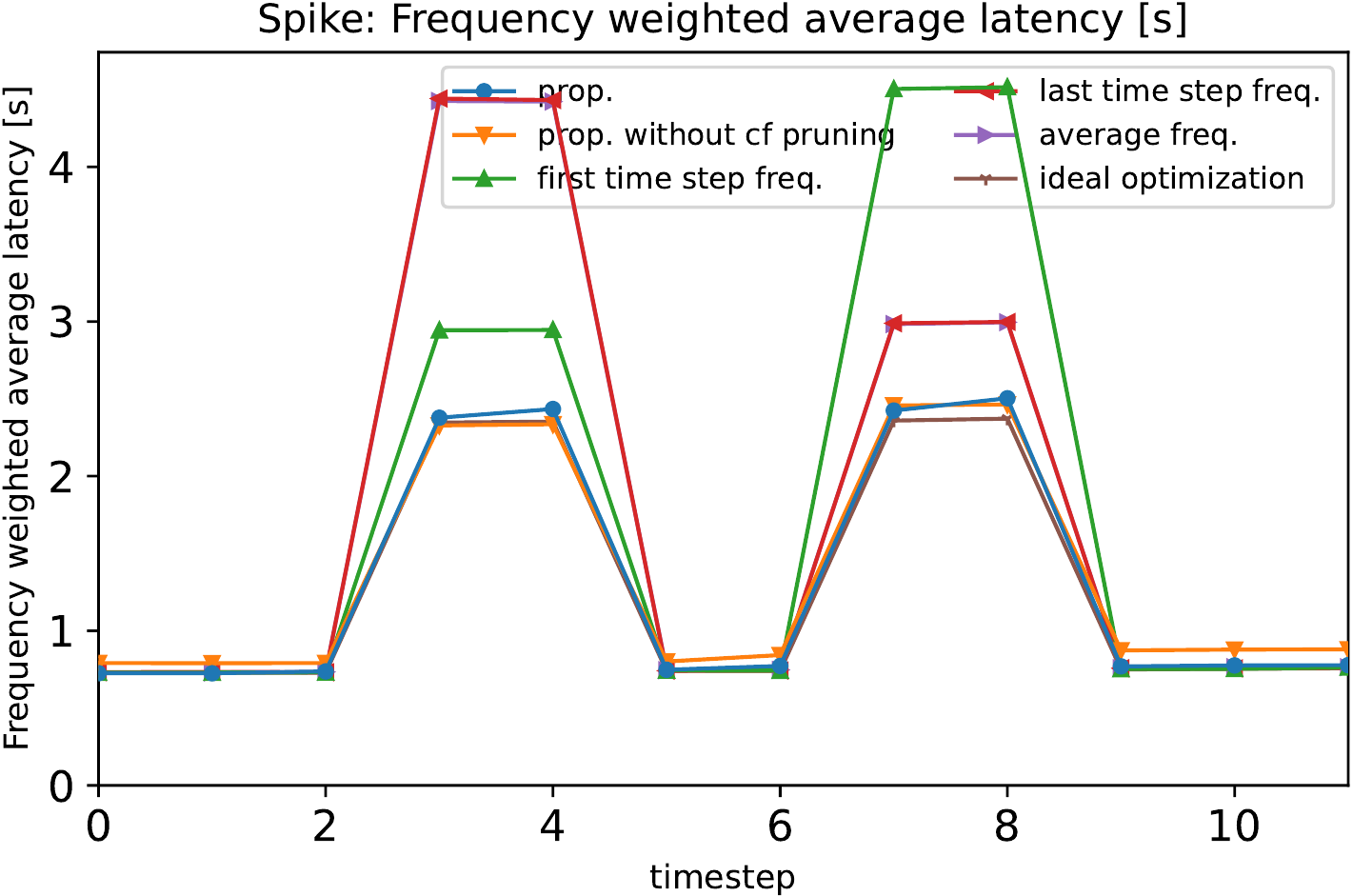}
    \caption{spike, 80per．この実験結果は加重平均の計算方法に修正が必要なため，参考図．修正した場合，時刻0,1,2,5,6,9,10 の全て最適化結果の応答時間が5倍で計算される見込み}
    \label{eval:spike_latency}
\end{figure}

\begin{figure}[t]
    \includegraphics[scale = 0.31, bb = 0 0 831 683]{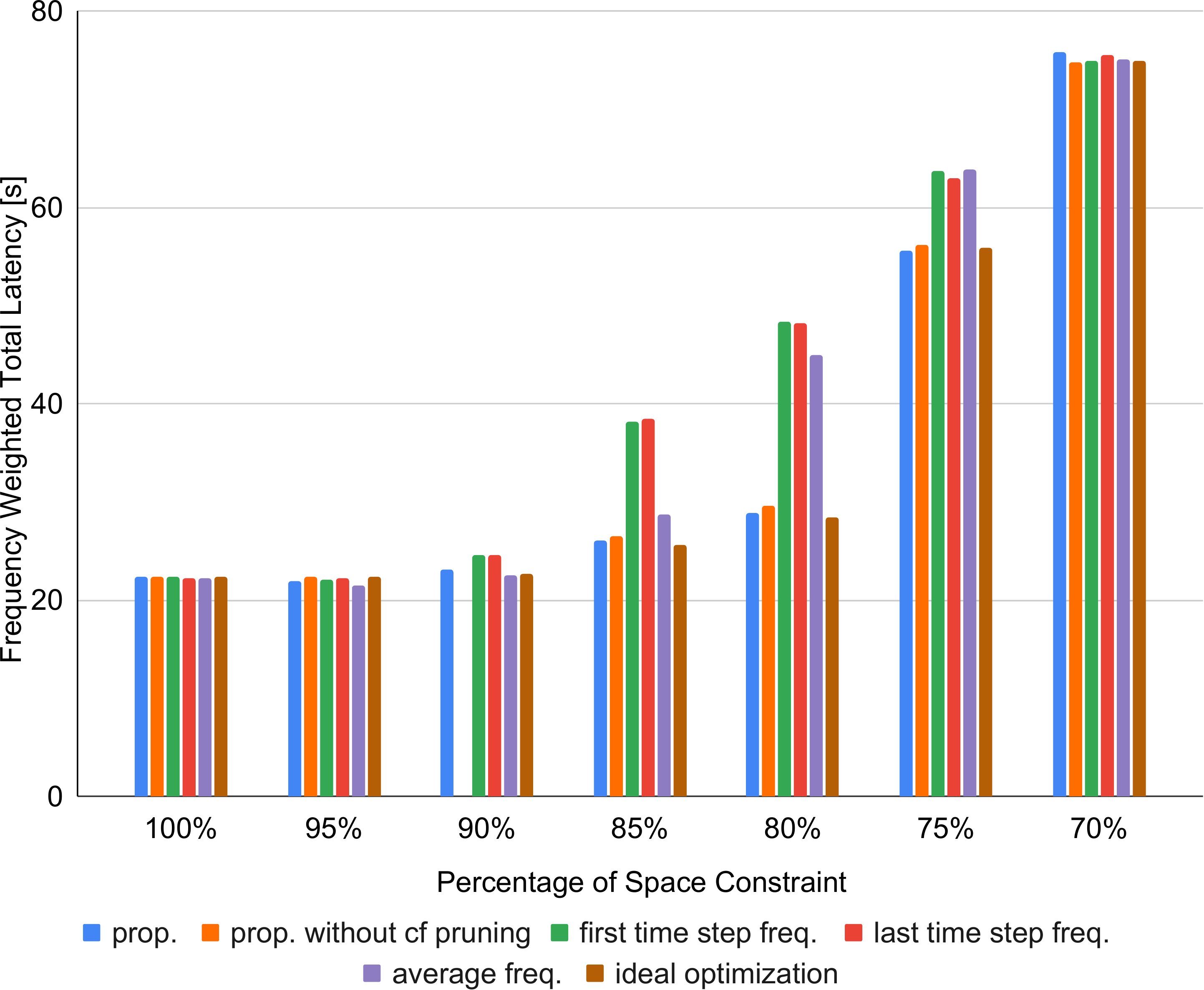}
    \caption{容量制約を変化させた際の各最適化手法の実行頻度による加重応答時間平均の総和の推移}
    \label{eval:latencies_under_various_space}
\end{figure}

\subsubsection{実行頻度が単調に変化するワークロードでの実験}

\subsubsection{実行頻度にピークを持つワークロードでの実験}

\todo{high priority}

\subsection{容量制約を変化させた際の応答時間の評価}

提案手法がマイグレーションを適切に推薦することで，容量制約の厳しい環境下でも応答時間の増加を低減することを確認するために，容量制約の値を変化させ実行頻度による加重応答時間平均の総和を計測した．
ただし，容量制約90\%での候補削減無しの提案手法は 24 時間最適化を実行しても最適化が完了しなかったため，本実験では対象外とした．
実験結果を図~\ref{eval:latencies_under_various_space} に示す．
ワークロードは実験~\ref{sec:periodical_latency}と同様に Periodical な頻度変化をする 12 時刻のワークロードを用いた．
また，平均実行頻度に対する最適化を容量制約無しで実行した最適化結果のストレージサイズを 100\% の値として用いた．
そのため，容量制約が100\% の環境が最も制約が緩く，容量制約が70\% の環境が最も制約が厳しい．
一般に，容量制約が厳しい場合ではストレージサイズを削減するためにスキーマを正規化する必要がある．
そして，スキーマを正規化すると，各クエリの実行計画としてジョインプランを選択する場合が増加するため，容量制約が厳しいほど応答時間が増加する．
ここで，提案手法ではマイグレーションを活用することで，各時刻で実行頻度の高いクエリに応答時間の短いクエリプランを推薦し，実行頻度による応答時間の加重平均の総和を削減する．
図~\ref{eval:latencies_under_various_space} の実験結果から，最小時刻実行頻度に対する最適化と最大時刻実行頻度に対する最適化の応答時間が容量制約85\%の場合に容量制約 90\% の場合に比べて大幅に増加している一方で、提案手法は応答時間の増加幅を低減していることを確認した．
また，容量制約80\% の場合では，平均実行頻度に対する最適化の応答時間が容量制約85\%の場合と比べて大幅に増加している一方で，提案手法は応答時間の増加を低減していることを確認した．
ただし，容量制約 75\% の場合では提案手法でも大幅な応答時間の増加を確認した．
これは，容量制約が厳しい場合では各クエリに対して提案可能な MV プランの総数が減少し，提案手法でもジョインプランを推薦する場合が増加するためと考えられる．
同様に，容量制約 70\% の場合では容量制約が厳しいため，提案手法はマイグレーションプランを推薦できず，他の手法と同等の応答時間となっている．

図~\ref{eval:latencies_under_various_space} において，コストモデルの精度が不十分なことが原因と考えられる3つの傾向が見られた．
これらの傾向の原因は，コストモデルの推定する実行コストと実際の応答時間の差異と考えられる．
1つ目は，理想的な最適化において，容量制約 90\% 下での実行頻度による加重平均応答時間が容量制約 95\% 下の実行頻度による加重平均応答時間に比べて小さい点である．
これは，容量制約が厳しくなるほど応答時間が増加する想定と異なる．
ここで，最適化の目的関数~\eqref{baseObjective:Objective} の最適化結果の値では，容量制約90\%の目的関数の値は，容量制約95\%の目的関数の値に対して，3.45\% 増加していた．
したがって，コストの最適化の結果では，容量制約が厳しくなるほど推定される実行コストは増加しており，想定と一致している．
2つ目は，提案手法と候補削減無しの提案手法の応答時間を比較すると，容量制約85\%, 80\%, 75\% において提案手法の応答時間が僅かに小さい点である．
これは提案手法が候補削減無しの提案手法に比べて限定された候補のみに対して最適化していることと矛盾する．
ここで，目的関数~\eqref{baseObjective:Objective} の値を比較すると，85\%, 80\%, 75\% の容量制約において目的関数の値の差異は 0.001\% 未満であった．
したがって，コスト最適化の結果では，提案手法と候補削減無しの提案手法の差異は僅かである．
3つ目は，容量制約 70\% の場合に提案手法の応答時間が最も大きくなっている点である．
ここで，目的関数~\eqref{baseObjective:Objective}の値を比較すると，各比較手法との目的関数の値の差は 0.1\% 未満であった．

提案手法は属性のカーディナリティや属性のフィールドサイズに基づくシンプルなコストモデルを使用している．
しかし，このコストモデルの精度が不十分であるため，目的関数の値と実際の応答時間の傾向に差異が生じたと考えられる．
各属性のカーディナリティやレコードサイズのより高精度な推定は今後の課題である．

\subsection{抽象化による候補削減による最適化実行時間の評価}\label{sec:evalOptRunningTime}

\begin{figure*}[t]
    \includegraphics[scale = 0.4, bb = 0 0 1234 521]{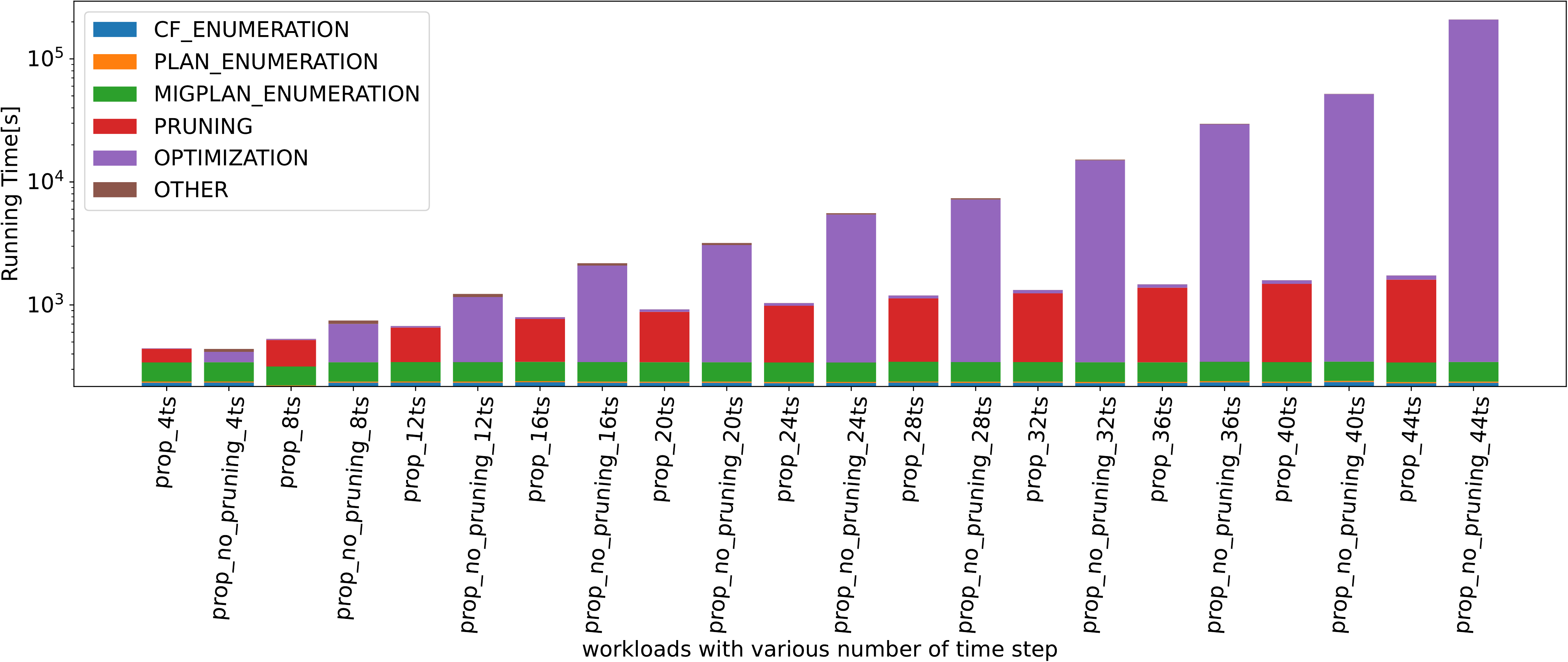}
    \caption{\todo{系列の名前を with pruning, without pruning に変更する．このキャプションで全て提案手法の実行時間を計測したものであることを明記する．また時刻毎に棒グラフを寄せて，他の時刻との間にスペースを追加することで，グラフを見やすくする}}
    \label{eval:latencies}
\end{figure*}

%\subsection{CF 列挙効率化による実行時間の評価}\label{sec:evalCFEnumerationTime}

%% file: parts/6_RelatedWork.tex
We summarize research trends in schema design for query workload on relational databases and NoSQL databases.

Since relational databases and NoSQL databases have different characteristics, we categorize existing methods designed for relational and NoSQL databases.

\paragraph{Schema design on relational databases}
Schema design methods for time-depended workload were 
proposed~\cite{pavlo17, Jindal2018, CloudViews2018, Peregrine2019, Kossmann2019}.
Pavlo et al.~\cite{pavlo17} proposed Peloton, a self-driving database framework for in-memory databases.
Peloton introduces recending-horizon control model (RHCM) to predict workload changes and also proposes a method that adaptively changes schema according to the workload changes. 
Kossmann et al.~\cite{Kossmann2019} proposed a dynamic optimization method for self-managing database systems.
Their method uses linear programming to determine an efficient order to tune multiple dependent features, such as index selection, compression schemes, and data placement.

In addition, there are other works ~\cite{Wiese2008, VanAken2017} that dynamically change various tuning parameters in database systems. 
In Wiese et al.~\cite{Wiese2008}, the database administrator registers a tuning procedure, and the  procedure is automatically triggered when a given condition is met, such as when deadlocks occur more frequently than a given threshold.
Aken et al.~\cite{VanAken2017} proposed OtterTune, which automatically tunes the memory size, cache size, etc. by leveraging past experiences: it combines supervised and unsupervised learning methods to choose the most impactful tuning knobs. 

However, those methods are difficult to be applied to NoSQL databases, because there is no clear distinction between physical schema and logical schema in NoSQL databases.

\paragraph{Schema design on NoSQL databases}
There are also studies on schema migration in NoSQL databases.
NoSE~\cite{Mior2017} was proposed as a NoSQL schema design method for static workloads.
NoSE estimates the execution cost of static workload query, and optimizes the schema design.
However, since it does not support time-depended workloads, even if the schema is optimized once, the performance may deteriorate due to workload changes.

There are studies on schema migration for time-depended workloads~\cite{Hillenbrand2020, boncz2019, CONST2020}.
Hillenbrand et al.~\cite{Hillenbrand2020} proposed a method that enumerates multiple data migration patterns using an input schema migration, and then chooses the optimized data migration method based on a rule-based manner.

Google Napa~\cite{DBLP:journals/pvldb/AgiwalLMRSZZCCD21}
guarantees robust query performance. Clients expect low query latency and low variance in latency regardless of the query/data ingestion load.
It also provides users a flexibility to tune the system and meet their goals based on data freshness, resource costs, and query performance.